\documentclass{elsart}
\usepackage{epsfig,graphicx, amssymb}
\begin{document}
\runauthor{Litak et al.}

\begin{frontmatter}
\title{Transition to Chaos  in the Self-Excited System with 
a Cubic Double Well Potential and Parametric Forcing}   
\author[Lublin1]{Grzegorz Litak\thanksref{E-mail},}
\author[Lublin1]{Marek Borowiec,}
\author[Lublin2]{Arkadiusz Syta,}
\author[Lublin1]{Kazimierz Szabelski}

\address[Lublin1]{Department of Applied Mechanics, Technical University of
Lublin,
Nadbystrzycka 36, PL-20-618 Lublin, Poland}
\address[Lublin2]{Department of Applied Mathematics, Technical University 
of
Lublin,
Nadbystrzycka 36, PL-20-618 Lublin, Poland}

\begin{abstract}
We examine the Melnikov criterion for a global homoclinic bifurcation 
and a possible transition to chaos in
case of
a single degree of freedom nonlinear oscillator with a symmetric 
double well nonlinear
potential.
The system was  subjected simultaneously to 
parametric periodic forcing and self excitation via negative damping 
term.
Detailed numerical studies confirm the analytical predictions and 
show that  transitions from regular to 
chaotic types of motion are often 
associated 
with
increasing the energy of an oscillator  and its escape from a single well. 
\end{abstract}

\thanks[E-mail]{Fax: +48-815250808; E-mail:
g.litak@pollub.pl (G. Litak)}

\end{frontmatter}

\section{Introduction}

A system of nonlinear stiffness having a square displacement force term,
and simultaneously,
 excited externally 
or 
parametrically   has been a subject
of studies for many years 
\cite{Szabelski1985,Szabelski1991,Litak1998,Rand2003,Rusinek2000,Rega1995,Szemplinska1993,Szemplinska1995,Lenci2003,Lenci2004}.
It has found numerous applications in mechanical 
engineering 
\cite{Szabelski1985,Szabelski1991,Litak1998,Rand2003,Rusinek2000,Rega1995,Szemplinska1993,Szemplinska1995} 
and 
control theory \cite{Lenci2003,Lenci2004,Tchoukuegno2003,Cao2005}. 
It was also one of intriguing 
examples 
of simple 
nonlinear systems which showed complex behaviour including chaotic oscillations \cite{Thompson1989}.  
Another important, partially separated, class   
of systems can be determined on the basis of nonlinear damping.
Such damping can, in some systems, change the sign depending on velocity or displacement values, and provide 
excitation energy to the examined system. 
These, so called, self--exited damping  
terms are
often used to describe systems with dry friction, bearings
lubricated by thin layer of oil, shimming in vehicle wheels or chatter in
cutting process,
\cite{Szabelski1985,Szabelski1991,Litak1998,Kapitaniak1991,Warminski2000}.
More recently such a nonlinear damping force has also been considered \cite{Li2004} 
in modelling of a 
modern 
vehicle 
suspension system due to electro- or magneto-rheological fluid damping
where it is causing a hysteretic effect. In this model \cite{Li2004} the authors used
a self-excited term of Rayleigh type and Duffing double well potential.

This class of a system has also been  investigated by Siewe {\em at al.} 
\cite{Siewe2004,Siewe2005}
in their works on   $\Phi^6$-Van der Pol oscillators. In  the first paper \cite{Siewe2004} they 
applied 
Melnikov approach and performed numerical simulations in case of  
 a potential based on polynomials of even orders (4th and 6th orders: $\Phi^4$ and 
$\Phi^6$ type 
potentials respectively)  
 and an external excitation. On the other hand in second paper \cite{Siewe2005} the 
triple well 
potential
based on polynomial of the  6th order was  studied in details for a dynamical system with 
external and parametric excitations. The Melnikov approach 
has been also 
considered 
in   
systems with a single well potential and various nonlinear damping terms.  
Among a large number of papers conducted that research we report 
\cite{Litak1999,Trueba2000,Awrejcewicz1999}. 
Litak {\em et al.} \cite{Litak1999} 
included the Rayleigh self-excited term to describe dynamics of the Froude 
pendulum,   Trueba {\em et al.}  \cite{Trueba2000} considered 
nonlinear damping 
following a power law in velocity ($\alpha (\dot x)^n$, $n=2,3$) for various systems, 
and finally  Awrejcewicz and Holicke \cite{Awrejcewicz1999} examined 
the effect of dry friction 
in a stick--slip system with the Duffing potential.

\begin{figure}[htb]
\centerline{
\epsfig{file=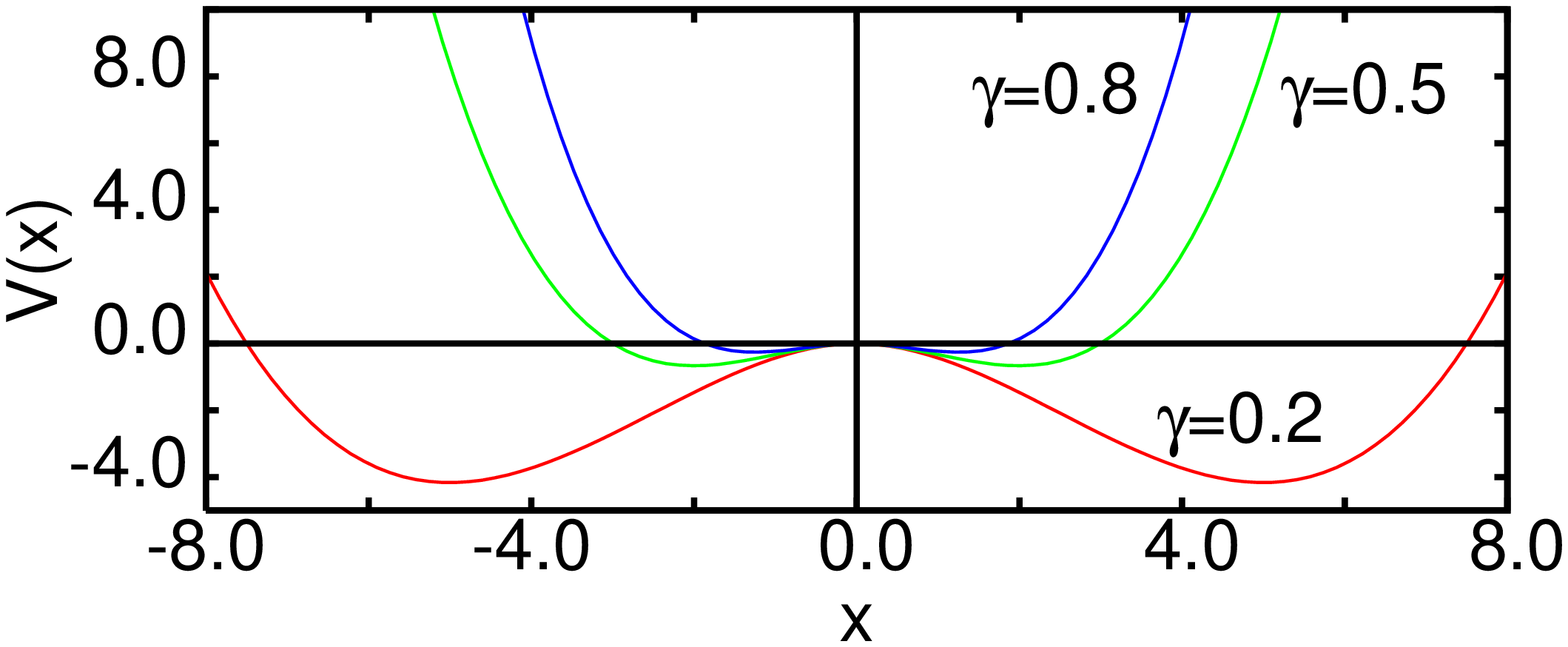,width=4.5cm,angle=-90}}
{\small \center Fig. 1. The double well potential  $V(x)=\delta x^2/2+\gamma |x|x^2/3$ for three
values of
$\gamma$ ($\gamma=0.8$, 0.5, and 0.2) and $\delta=-1.0$ .}
\end{figure}

 Note, that all above systems can be still regarded as simple ones but its 
combined 
nonlinear answer on  
excitation is complicated and deserves detailed investigations.

In this note we shall examine  
transition from regular oscillations to chaos in a simple, one degree of 
freedom, system subjected 
to parametric and self excitations
with a square albeit symmetric
stiffness:
\begin{equation}
\label{eq1}
\ddot{x} + \alpha (-1+x^2) \dot{x}
+ (\delta- \mu \cos{ 2 \omega t}) x +\gamma |x|x=0,
\end{equation}
where $x$ is a displacement  $\alpha (-1+x^2) \dot{x}$ is nonlinear 
damping,
$-\mu x\cos{ 2 \omega t}$ is a  parametric excitation while 
$\gamma |x|x$ and $\delta x$
are square and linear force terms. 
It is worth mentioning that, the corresponding potential can be described by two polynomials of the 3rd 
order ($\Phi^3$ type) for $x > 0$ and $x < 0$, respectively:
\begin{equation}
\label{eq2}
V(x)=\frac{\delta x^2}{2} + \frac{\gamma |x| x^2}{3}
\end{equation}
and plotted in Fig. 1
for $\gamma=0.2$, $\gamma=0.5$ and  $\gamma=0.8$.
In spite of appearing $|x|$ in Eq. \ref{eq2} the function $V(x)$  is of 
$C^2$ class at $x=0$ because it is approaching to 0, for 
$x \rightarrow 0$, as $\pm x^3$.

The present paper consists of   three sections and two appendixes.
After a short introduction in the present section (Sec. 1)
we will examine the above system analytically and numerically in Sec. 2.
There we will especially 
focus on a transition and a corresponding scenario from regular to chaotic vibrations.
In  Sec. 3 we will discuss the numerical simulations verifying the analytical findings.
In the last section (Sec. 4) we will provide the summary and conclusions. 
In the appendixes we will present different treatment of a self-excitation term and its consequences to 
the applied analytical methods (Appendix A) and show the Melnikov 
integration procedures in details (Appendix B).

\section{Melnikov Analysis}

We are starting our study from the second order equation of motion (Eq. 
\ref{eq1}).
After transforming it into two differential equations of the  first order, 
a standard procedure in 
examining 
homoclinic transition will be applied to
look for stable and unstable manifolds and their possible cross-sections in  presence of weak excitations and 
damping.  Therefore  we have introduced a small 
parameter
$\varepsilon$ to the above
equations enabling these terms to be switched on 
\cite{Rand2003,Guckenheimer1983,Wiggins1990}:

\begin{eqnarray}
\label{eq3}
\dot{x} &=& v \\
\dot{v}  &=& -\delta x - \gamma |x|x + \varepsilon \left[ \tilde{\alpha} (
1-x^2) v + x \tilde{\mu} \cos{\left(2 \omega t \right)}
\right], \nonumber
\end{eqnarray}

where $\tilde{\alpha} \varepsilon=\alpha$ and $\tilde{\mu} \varepsilon=\mu$.
Note, that this is not a unique way of splitting the initial differential 
equation 
  of the second 
order (Eq. \ref{eq1}) into two equations of the first order.

The other way is connected with 
different treatment of 
the self-excitation term.
Basing on 'fast' and 'slow' variables ($x,w$) identification
in Van der Pol 
fashion  
\cite{Strogatz1994} we can write: 

\begin{figure}[htb]
\vspace{0cm}

\epsfig{file=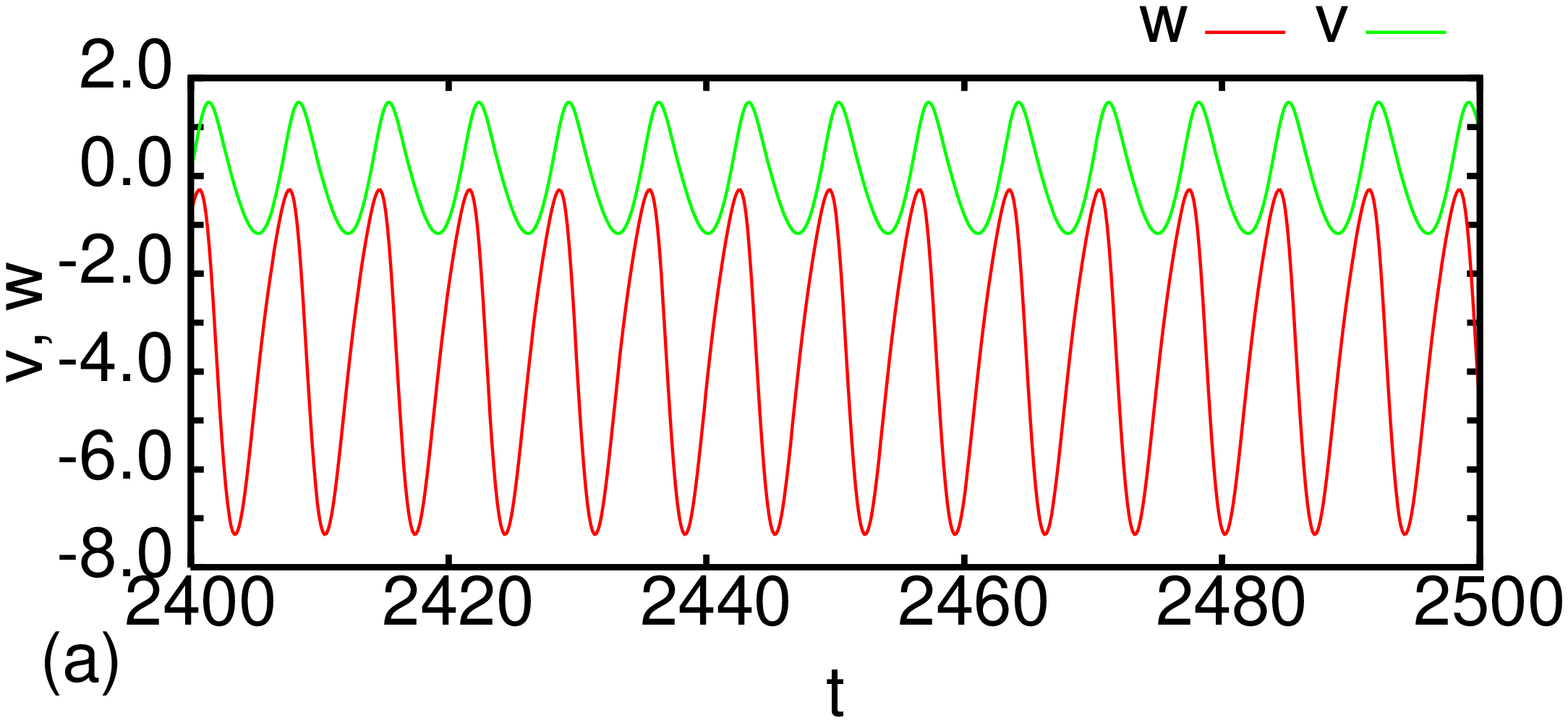,width=4.5cm,angle=-90}

\epsfig{file=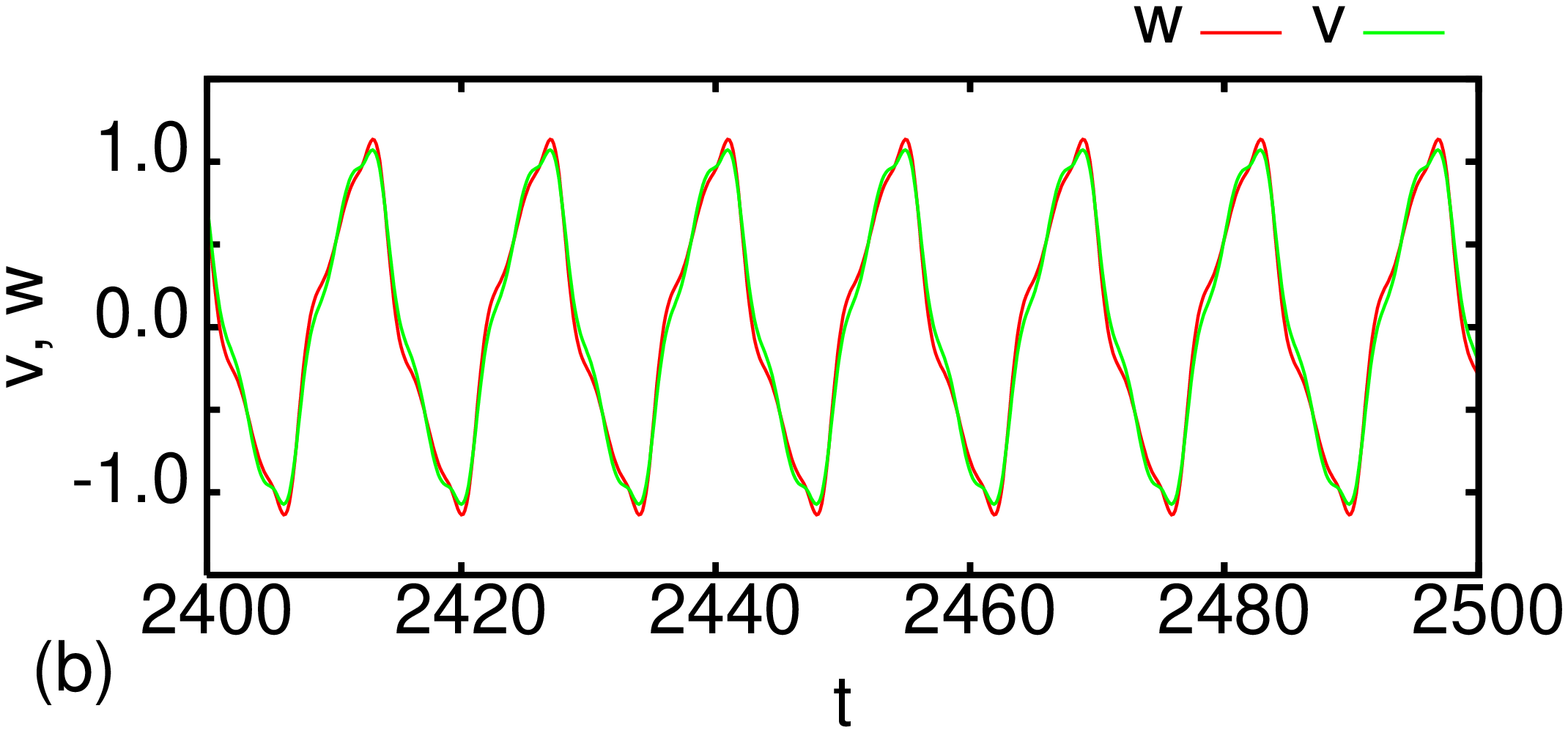,width=4.5cm,angle=-90}

\epsfig{file=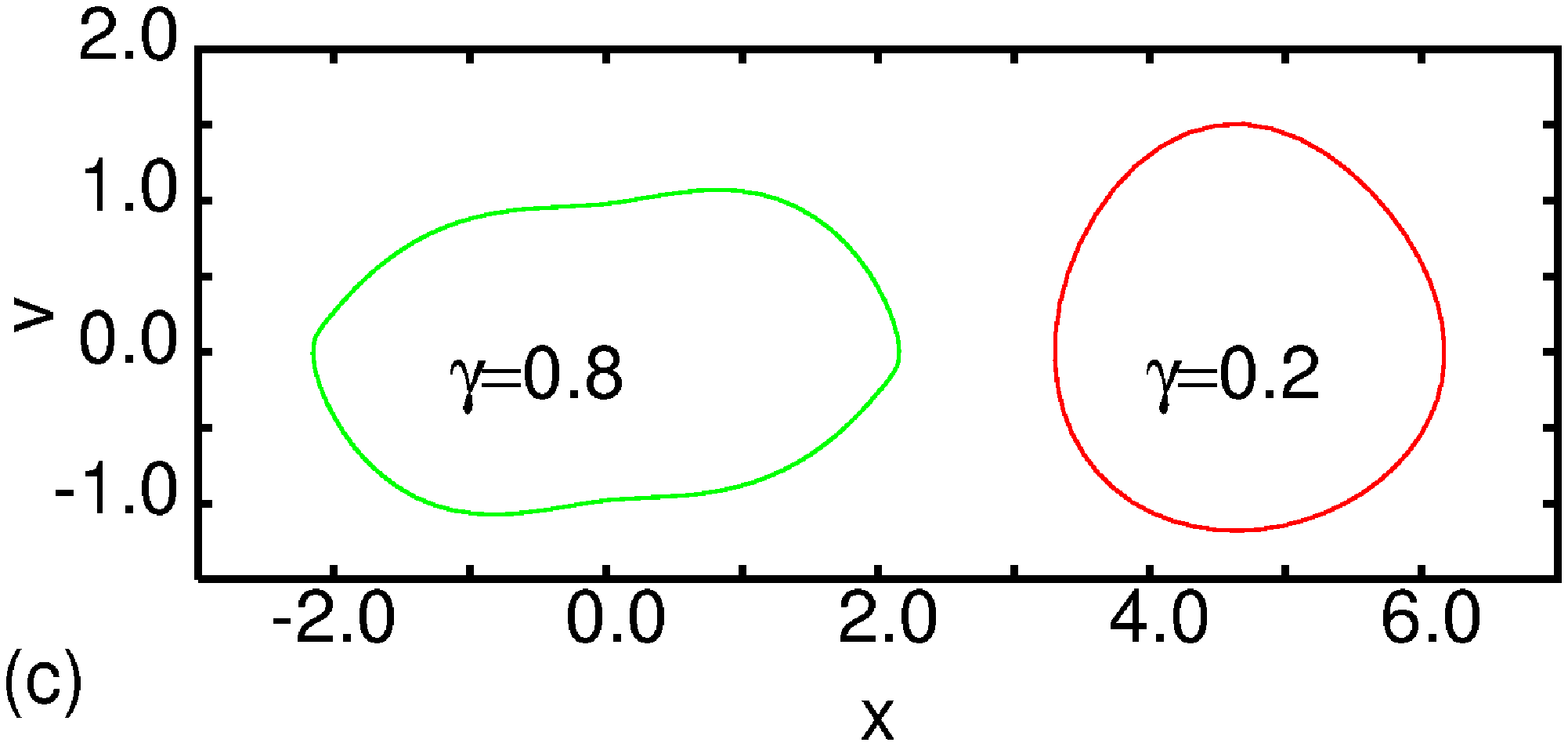,width=4.5cm,angle=-90}

{\small \center Fig. 2. \label{fig2} Comparison of $v$ and $w$ time 
histories for 
$\gamma=0.2$
(Fig. 2a) and
$\gamma=0.8$ (Fig. 2b). Other
system parameters: $\alpha=0.1$
$\omega=0.45$,$\delta=-1.0$, $\mu=0.6$. (Fig. 2c) Phase portraits (in 
$x$-$v$ space) of for both cases ($\gamma=0.2$ 
and 0.8).    
}
\end{figure}

\begin{equation}
\label{eq4}
w=\dot x+ \varepsilon \tilde{\alpha} \left(
x-\frac{x^3}{3}  \right). 
\end{equation}

This possibility has been examined thoroughly in  Appendix A.
Note, the difference between $w(t)$ and $v(t)$ depends on 
the influence of the self-excitation term leading to relaxation 
oscillations in the Van der Pol system. The time histories of 
$w$ and $v$ (Figs. 2a-b) show that this term could more influential for 
smaller $\gamma$ ($\gamma=0.2$, Fig. 2a) while a case with larger $\gamma$ 
($\gamma=0.8$, Fig. 1b)  is apparently negligible.
For better clarity we plotted phase portraits (Fig. 2c) for both cases showing that they
 correspond to steady state vibrations located in different regions of the phase plane. This is associated with 
differences in a potential shape (Fig. 1).

In the first case vibrations have been realized around the minimum of the potential $V(x)$ ($x \approx 
5.0$) 
inside one of  
its wells, while in the 
second one vibrations are located around $x=0$ between $x=-2$ and 2. As in the examined example the 
self-excitation Van der Pol term:
\begin{equation}
\label{eq5}
{\rm VdP}(x,\dot{x})=\alpha (-1+x^2) \dot{x}
\end{equation}
is changing  its sign at $x= \pm 1.0$. One can easily see that for 
$\gamma=0.2$ the system behaves if it 
possessed renormalised nonlinear damping term which does not change 
its sign. Moreover this term 
is positively defined. Thus the system, due to the shape of the external potential (Fig. 1),  does not have 
any relaxation character typical for simple Van der Pol oscillator and the distinction of 'slow' 
and 'fast'  variables is not relevant here. This fact has its reminiscence 
in a large difference in $v(t)$ and 
$w(t)$ time histories in Fig. 2a. The transformation $v(t) \rightarrow w(t)$   can be better 
argued in the 
the second examined case ($\gamma=0.8$). There the  relaxation vibrations are present but in 
this case 
the effect of self-excitation turns out be small (Fig. 2b).   
 
Note, that the 
unperturbed Hamiltonian 
$H^0$ is the same for both cases (with $v$ or $w$ variables) and reads:
\begin{equation}
\label{eq6}
H^0= \frac{v^2}{2} + V(x).
\end{equation}

The potential function $V(x)$ (Fig. 1) has the local peak  at 
$x_0=0$.
The existence of this point with a horizontal tangent makes  
global homoclinic bifurcations, including transition from 
regular to 
chaotic 
solution, which may occur in the system.
Obviously, at this point the system velocity reaches zero value ($v=0)$ (Fig. 1). Thus, 
according to our potential gauge,
the 
total energy has only its 
potential part which is  also zero:

\begin{equation}
\label{eq7}
E=V(x=0)=0.
\end{equation}

Transforming Eqs. \ref{eq4},\ref{eq6} for a constant energy, chosen here as 
zero, (Eq. \ref{eq7}) 
we obtain the following expression for velocity:

\begin{equation}
\label{eq8}
v= \frac{{\rm d} x}{{\rm d} t} =
\sqrt{2 \left(-  
\frac{\delta x^2}{2} - \frac{\gamma 
|x|x^2}{3}\right)},
\end{equation}

from which

\begin{equation}
\label{eq9}
t-t_0=\int \frac{1}{\sqrt{2 \left(-
\frac{\delta x^2}{2} - \frac{\gamma
|x|x^2}{3}\right)}}  {\rm d} x,
\end{equation}

where $t_0$ represents an integration constant.
As a result of integration (Eq. \ref{eq9}) we get so called homoclinic orbits (Fig. 3) 
parametrised by time $t$:

\begin{eqnarray}
x^*=x^*(t-t_0) &=& \pm \frac{3 \delta}{2 \gamma} \left( 1 -  
\tanh^2 \left( \frac{\sqrt{-\delta} (t-t_0)}{2}\right) \right) 
\nonumber \\
v^*=v^*(t-t_0) &=& \mp \frac{3 \delta \sqrt{-\delta}}{2 \gamma} \frac{\tanh \left( 
\frac{\sqrt{-\delta} (t-t_0)}{2}\right)}{ \cosh^2 \left( 
\frac{\sqrt{-\delta} (t-t_0)}{2}\right)},
\label{eq10}
\end{eqnarray}
where '$+$' and '$-$' signs are related to I and II orbits (Fig. 3), respectively. 
\begin{figure}[htb]
\vspace{-3cm}
\centerline{
\epsfig{file=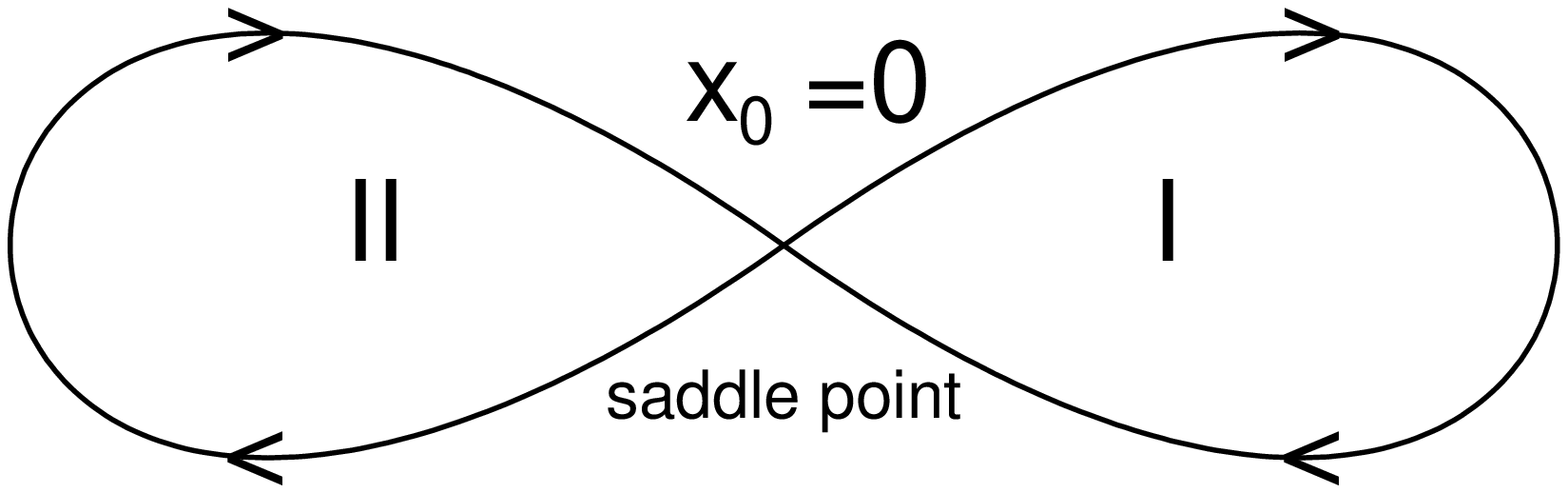,width=10.5cm,angle=-90}}
\vspace{-3cm}
~

{\small \center Fig. 3. \label{fig3} Homoclinic orbits (I and II)  of 
unperturbed potential
 $V(x)=-x^2/2+|x|x^2/3$ (for $\delta=-1$ and $\gamma=1$) in a phase plane 
$(x,\dot x)$.
For  $t \rightarrow \pm \infty$ we get
$(x,\dot x) \rightarrow (x_0,\dot x_0)=(0,0)$. 
}
\end{figure}

Note, the central saddle point $x_0=0$ is reached in time $t$ 
corresponding to $+\infty$ and $-\infty$, respectively.

In case of perturbed orbits $W^S$ (a stable manifold) and $W^U$ 
(an unstable manifold) the 
distance between them is given 
by 
the Melnikov function ${\rm M}(t_0)$ 
\cite{Guckenheimer1983,Wiggins1990,Tyrkiel2005}:

\begin{equation}
\label{eq11}
{\rm M}(t_0) = \int_{- \infty}^{ + \infty}  h( x^*, v^*)  \wedge g( x^*,
v^*) {\rm d} t
\end{equation}

where the corresponding differential forms $h$ is the gradient of unperturbed 
Hamiltonian (Eq. \ref{eq6}): 

\begin{equation}
\label{eq12}
h = \left(\delta x + \gamma |x^*|x^*\right) {\rm d} x  + v {\rm d}v, 
\end{equation}

and $g$ is a perturbation form (Eq. \ref{eq3}) to the same Hamiltonian (Eq. \ref{eq6}):

\begin{equation}
\label{eq13}
g = \left( \tilde{\mu} x \cos{2 \omega \tau} + \tilde{\alpha}
\left(1-x^2 \right)
v \right) {\rm d}x 
\end{equation}

Both differential forms are
defined on homoclinic manifold $(x,v)=(x^*,v^*)$ (Eq. \ref{eq10}, Fig. 2).

\begin{figure}[htb]
\centerline{
\epsfig{file=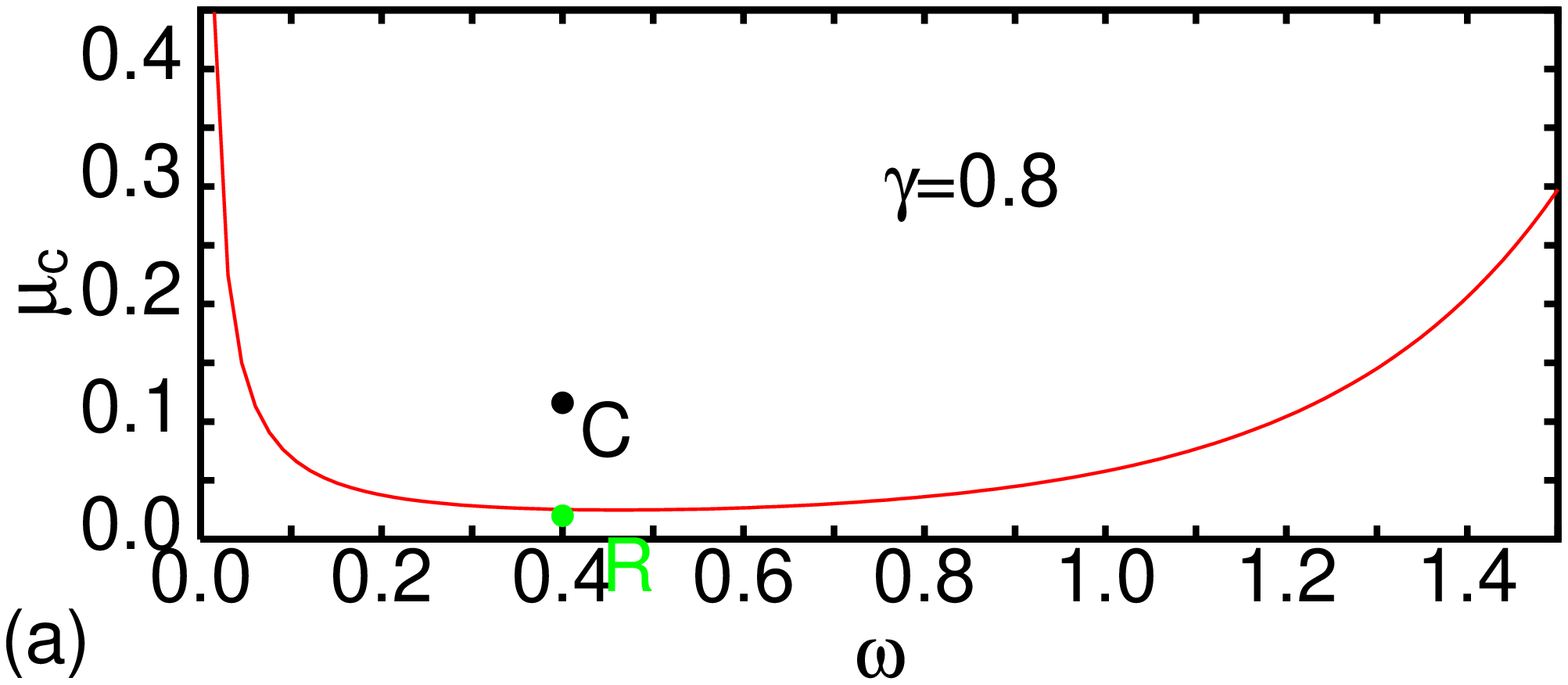,width=4.5cm,angle=-90}} 

\centerline{
\epsfig{file=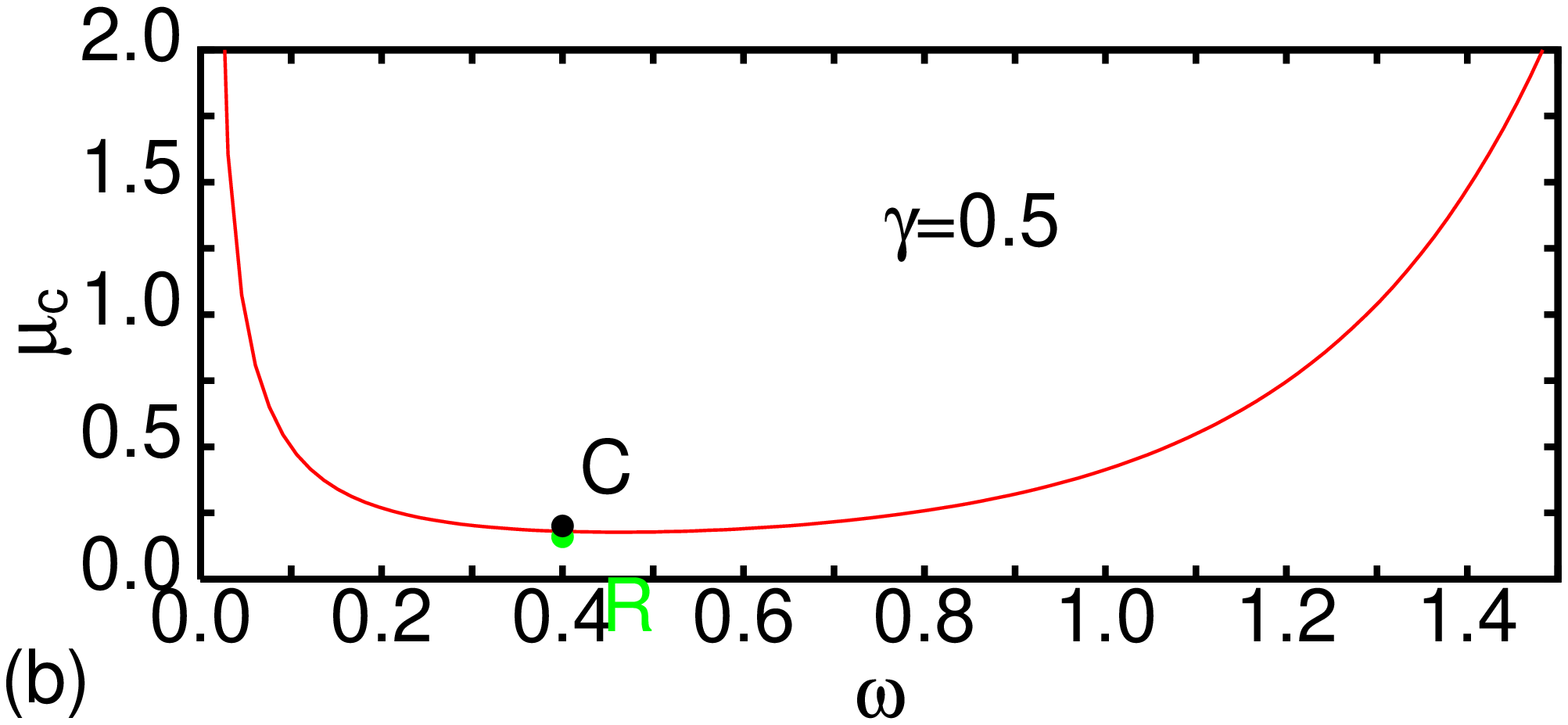,width=4.5cm,angle=-90}} 

\centerline{
\epsfig{file=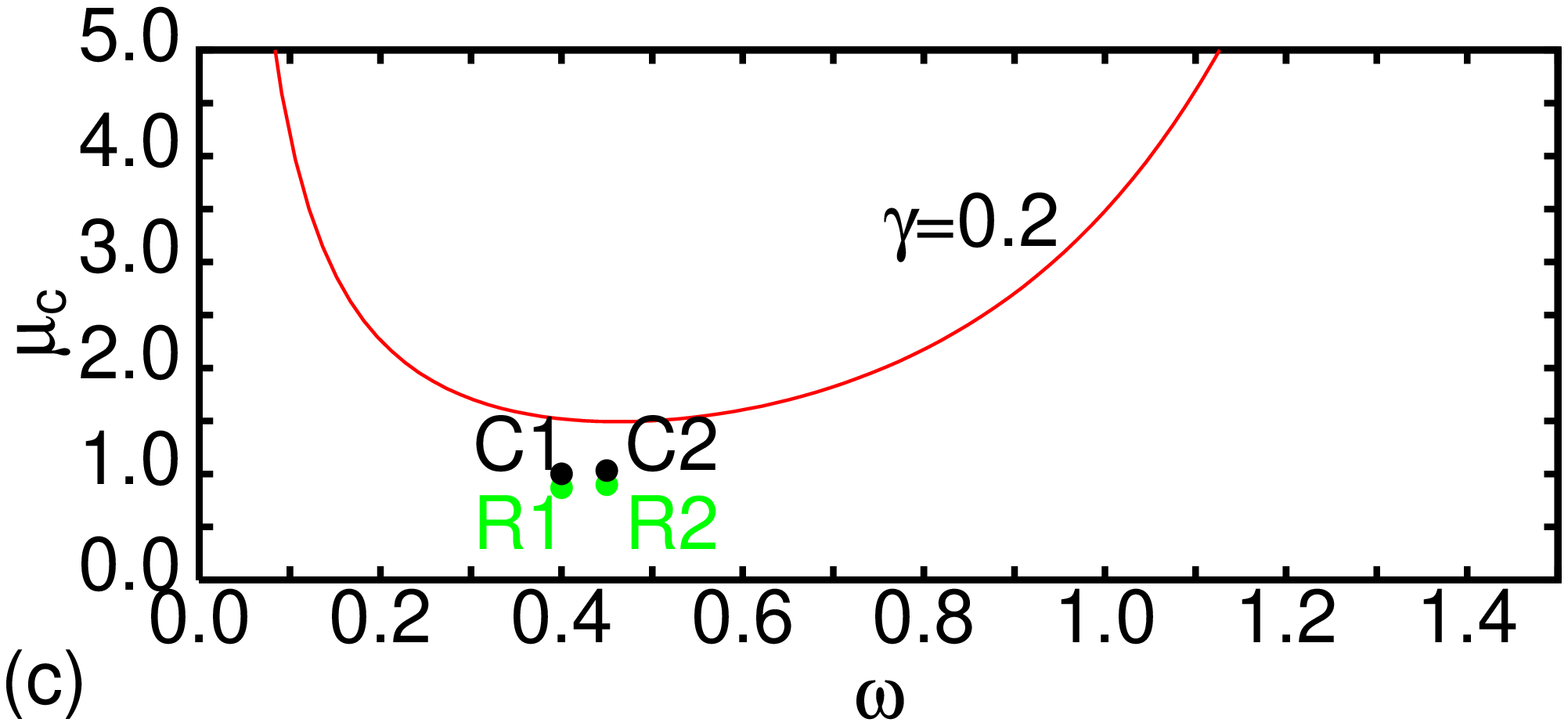,width=4.5cm,angle=-90}}

\centerline{
\epsfig{file=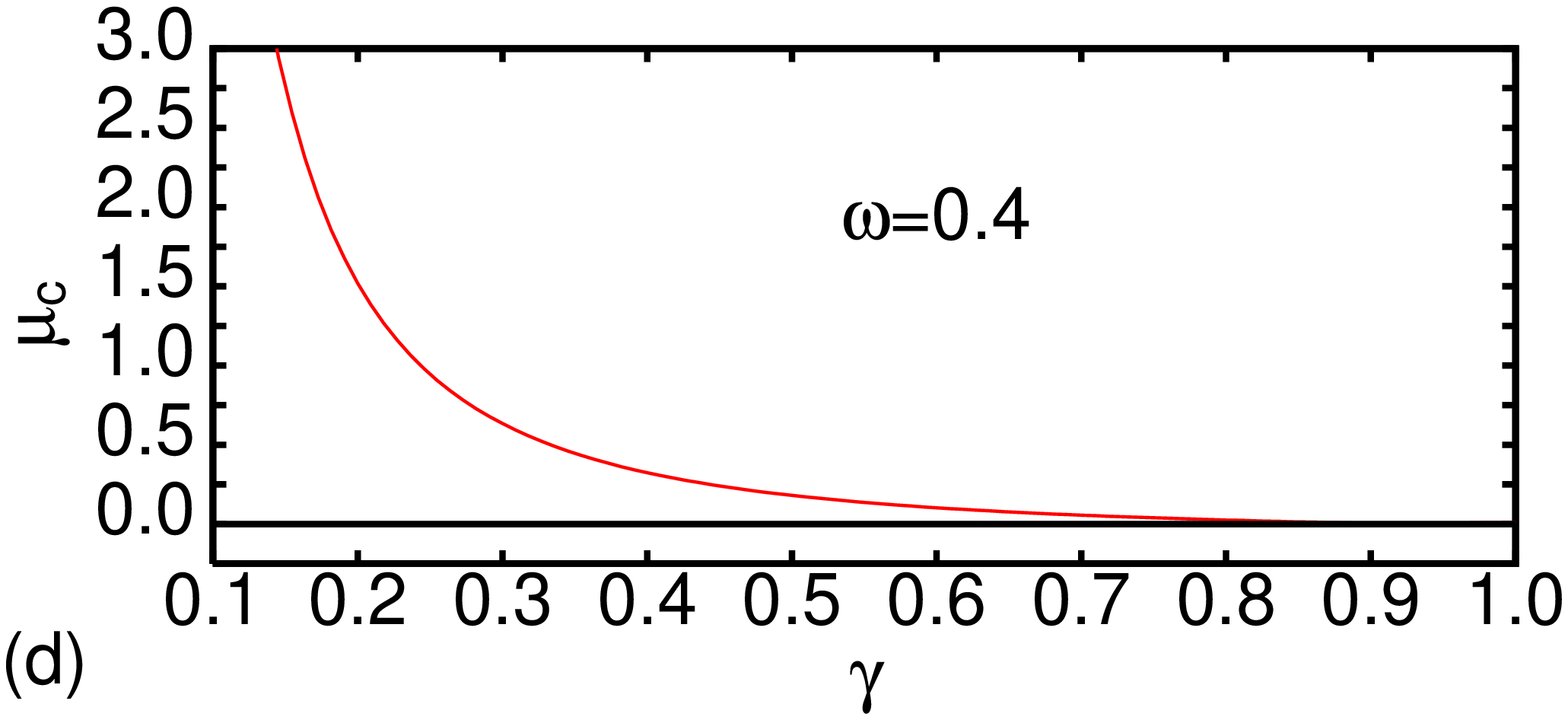,width=4.5cm,angle=-90}}

{\small \center Fig. 4. \label{fig4} 
Critical value of $\mu_c$, for $\alpha=0.1$, versus $\omega$ for $\delta=-1$
and various $\gamma$
($\gamma=0.8$ (Fig. 4a), $\gamma=0.5$ (Fig. 4b), $\gamma=0.2$ (Fig. 4c))
and versus $\gamma$ (Fig. 4d).
}
\end{figure}

\begin{figure}[htb]
\centerline{
\epsfig{file=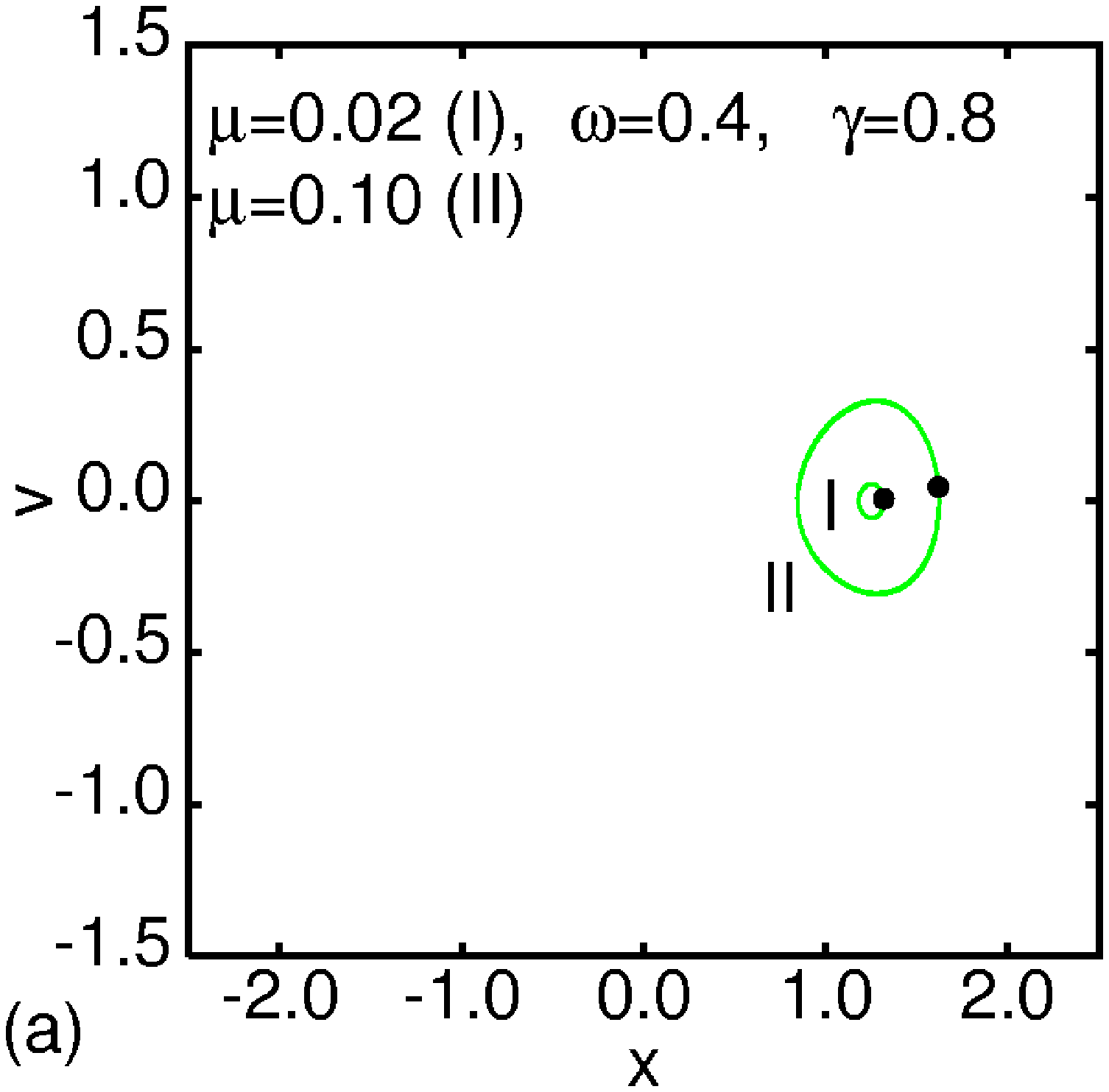,width=6.5cm,angle=-90}
\epsfig{file=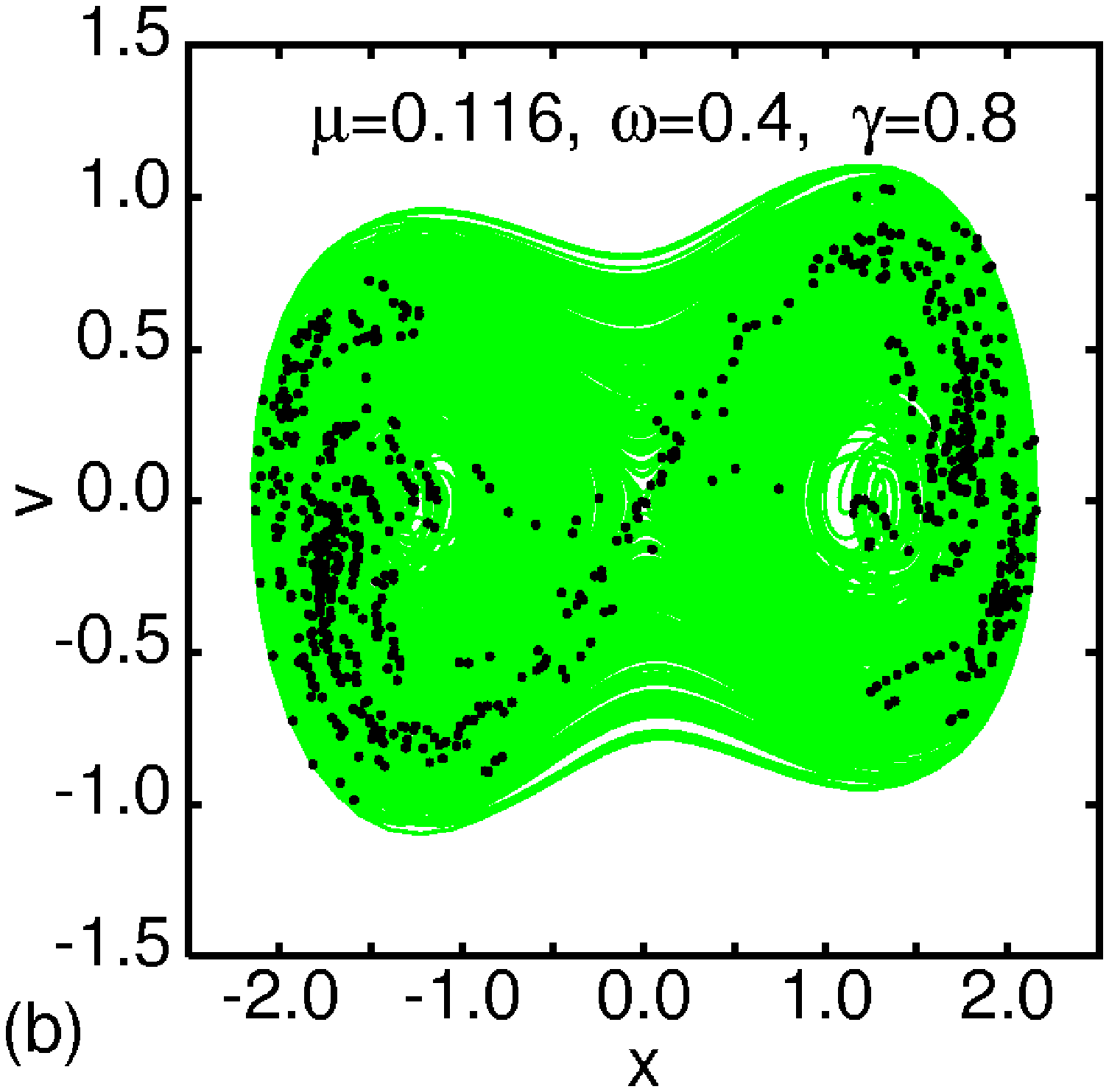,width=6.5cm,angle=-90}}
\vspace{-0.5cm}
~

\centerline{
\epsfig{file=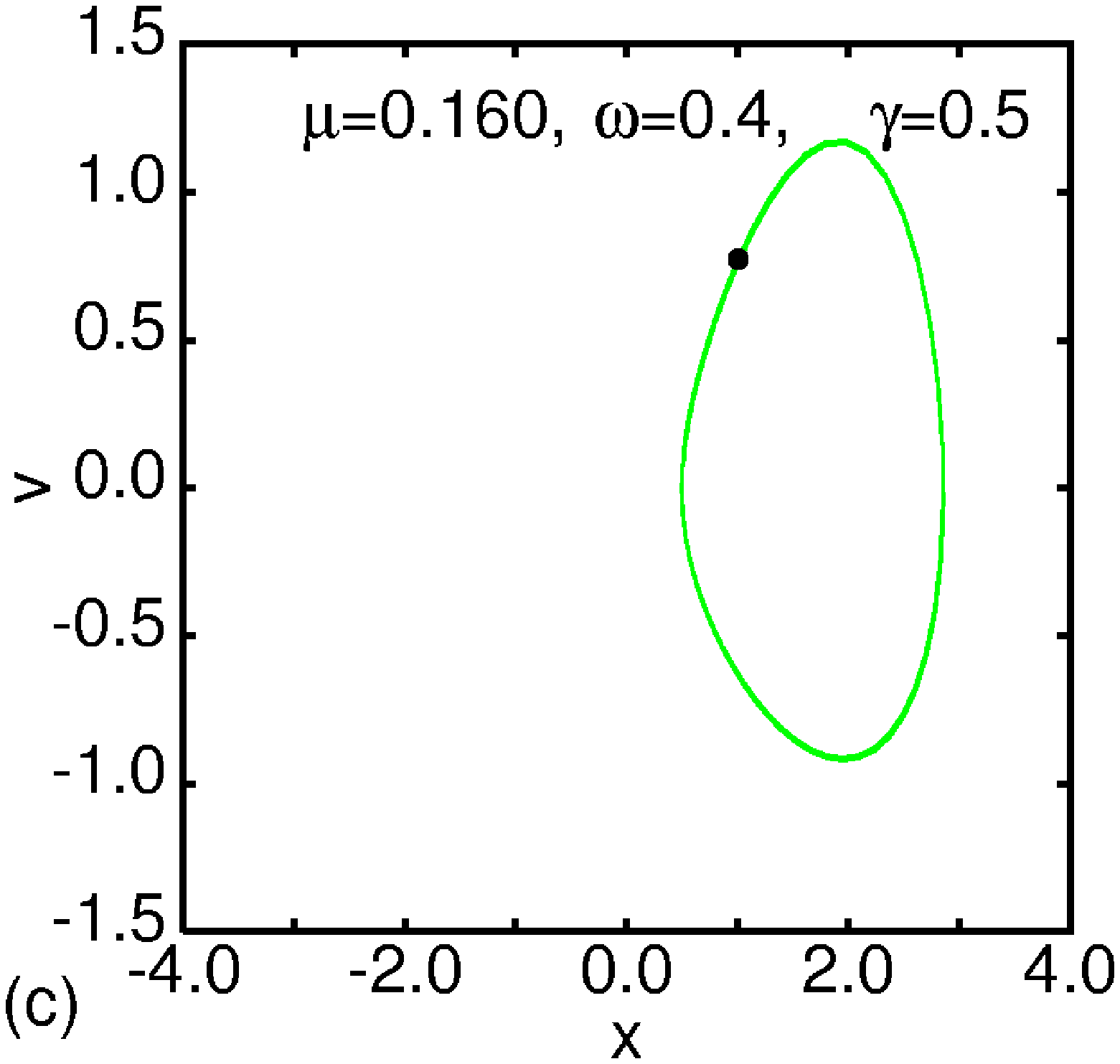,width=6.5cm,angle=-90}
\epsfig{file=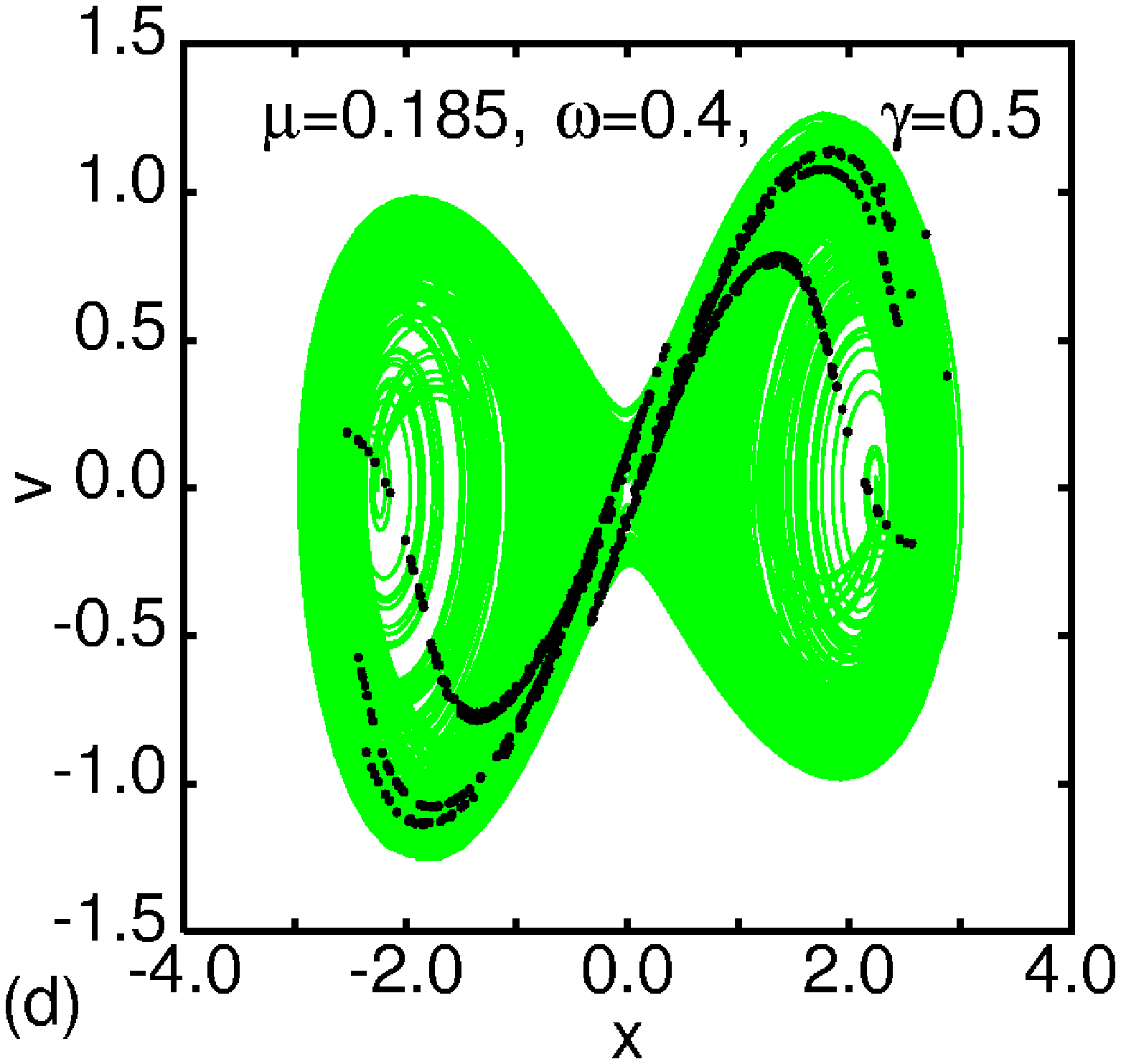,width=6.5cm,angle=-90}}

{\small \center Fig. 5. \label{fig5}
Phase diagrams and Poincare sections (simulations were based on Eq. 
\ref{eq1}) chosen sets system
parameters $\alpha=0.1$, $\delta=-1.0$ and $\mu$, 
$\omega$, $\gamma$ (corresponding values 
are indicated in figures) denoted by 
points in Figs. 4a-d.
The corresponding maximal Lyapunov exponent is (Fig. 5a): 
$\lambda_1=-0.0279$ 
for '1' and  
$\lambda_1=-0.0238$ for '2',
(Fig. 5b): $\lambda_1=0.0903$,
(Fig. 5c): $\lambda_1=-0.0978$), (Fig. 5d): $\lambda_1=0.0812$, (Fig. 5e):
$\lambda_1=-0.1513$, (Fig. 5f):
$\lambda_1=0.1202$, (Fig. 5g): $\lambda_1=-0.1974$, (Fig. 5h): 
$\lambda_1=0.0529$, respectively.}
\end{figure}

\begin{figure}
\centerline{\hspace{0.1cm}
\epsfig{file=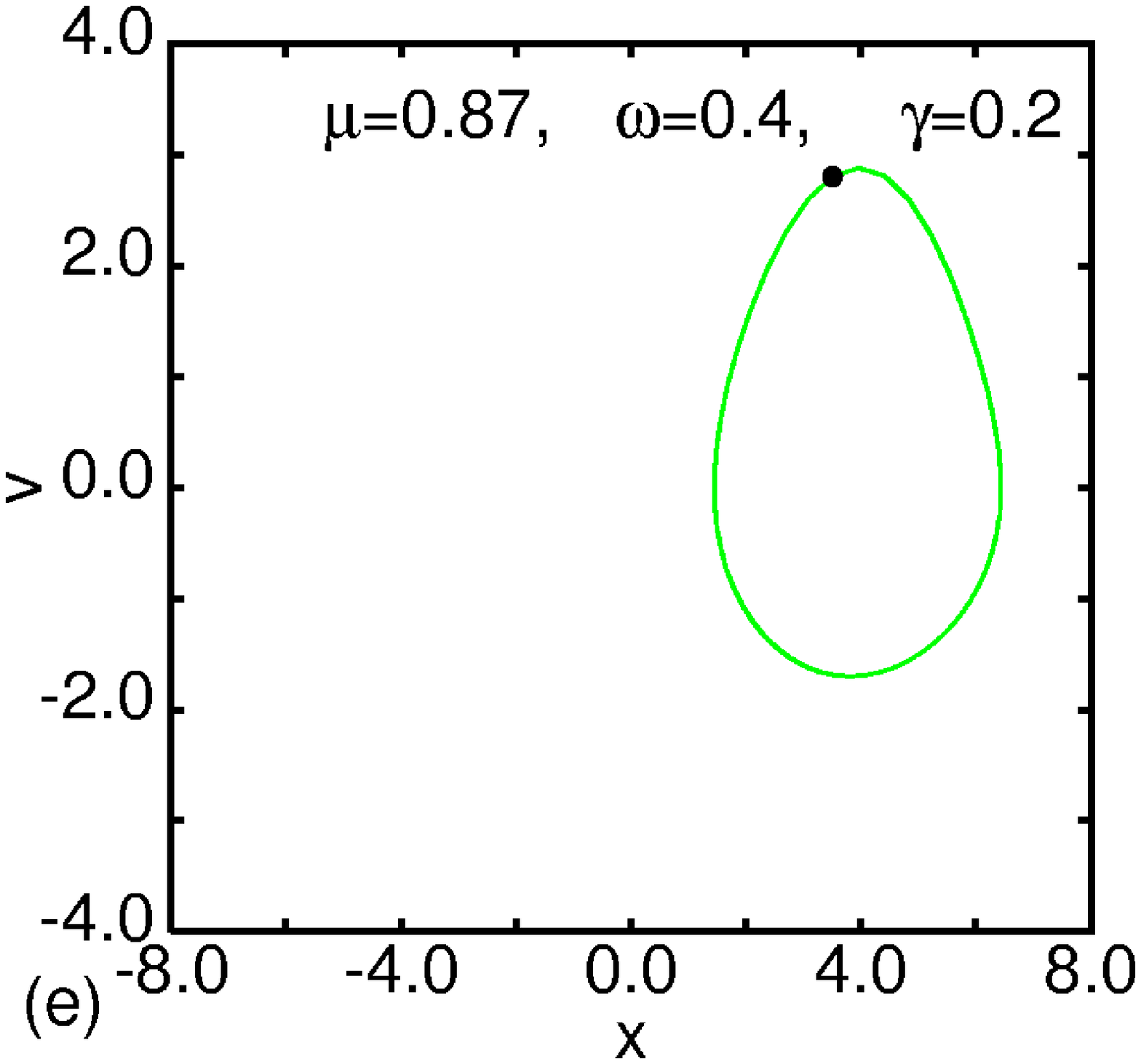,width=6.3cm,angle=-90} \hspace{0.1cm}
\epsfig{file=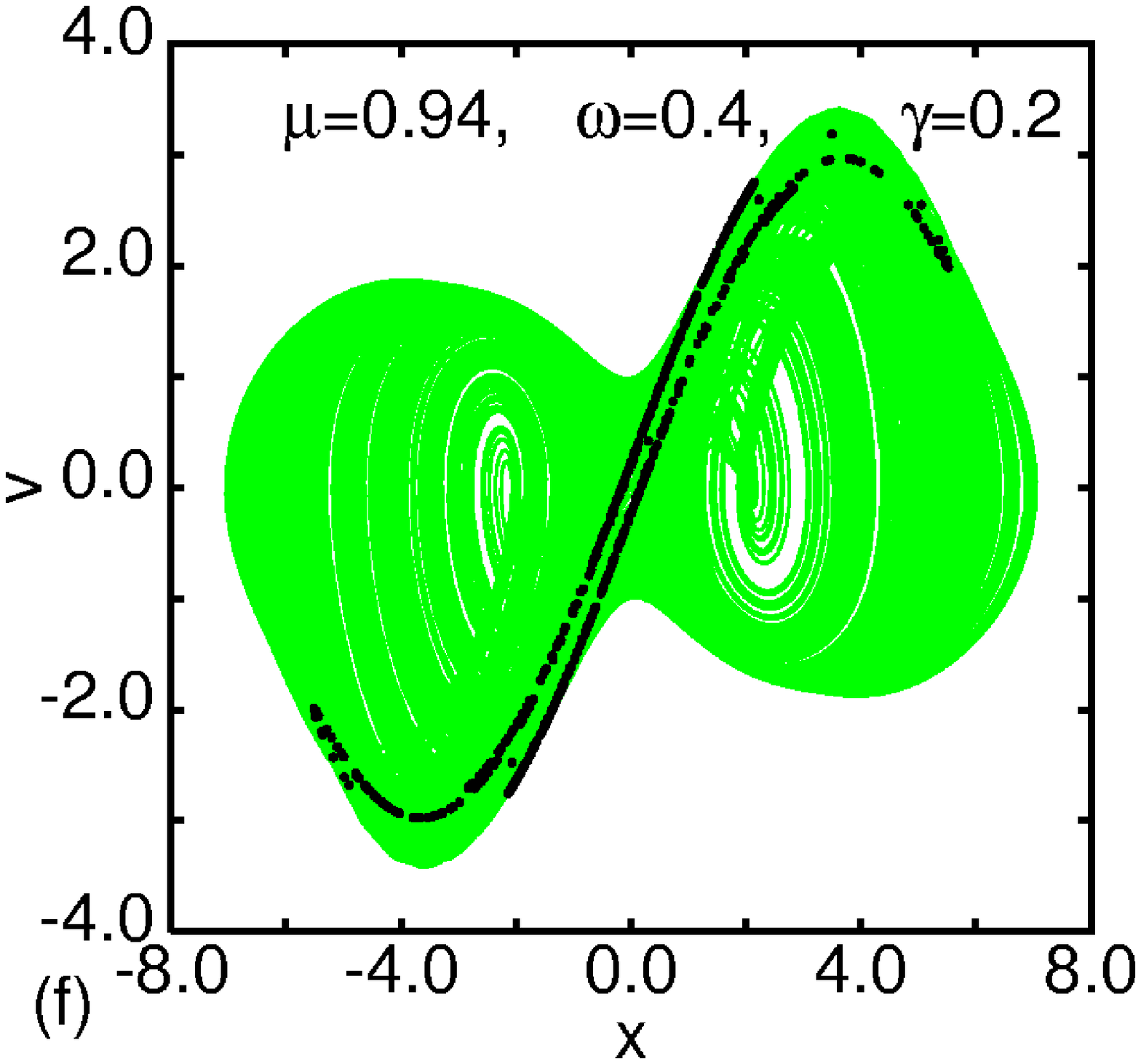,width=6.3cm,angle=-90}}
\vspace{-0.5cm}
~

\centerline{ \hspace{0.1cm}
\epsfig{file=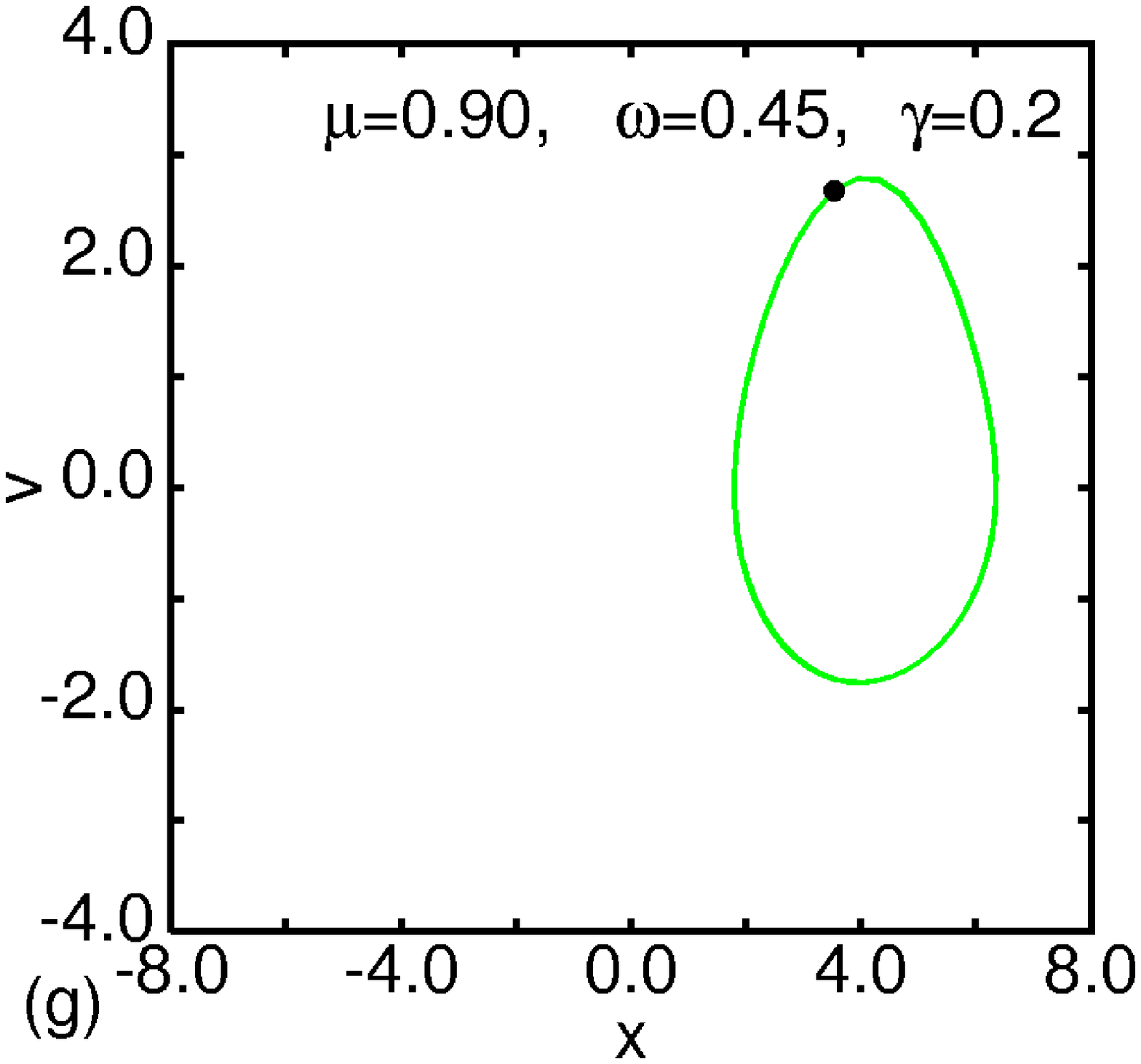,width=6.3cm,angle=-90} \hspace{0.1cm}
\epsfig{file=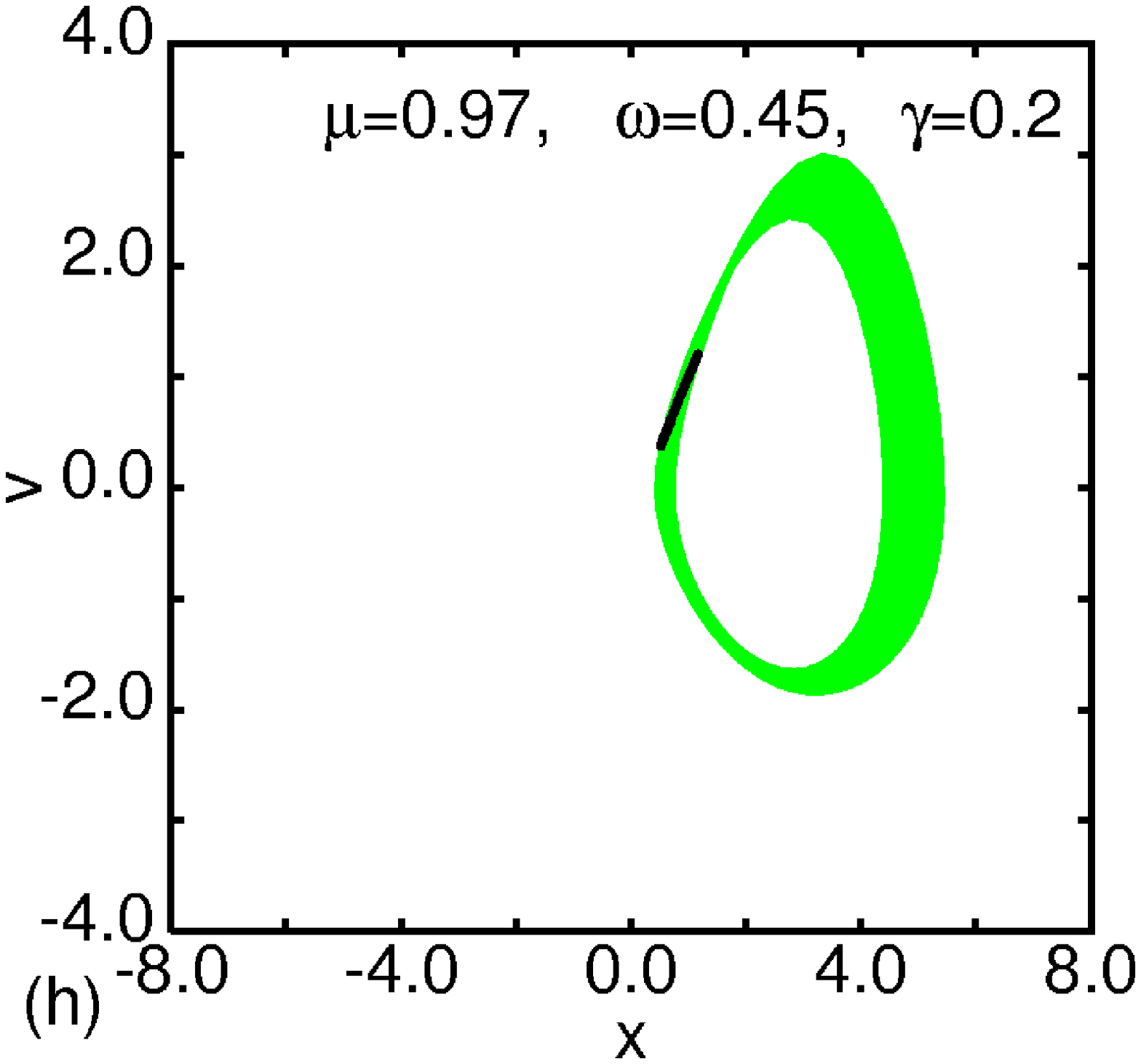,width=6.3cm,angle=-90}}

{\small \center Fig. 5. Continuation.
}
\end{figure}

\begin{figure}[htb]
\centerline{
\epsfig{file=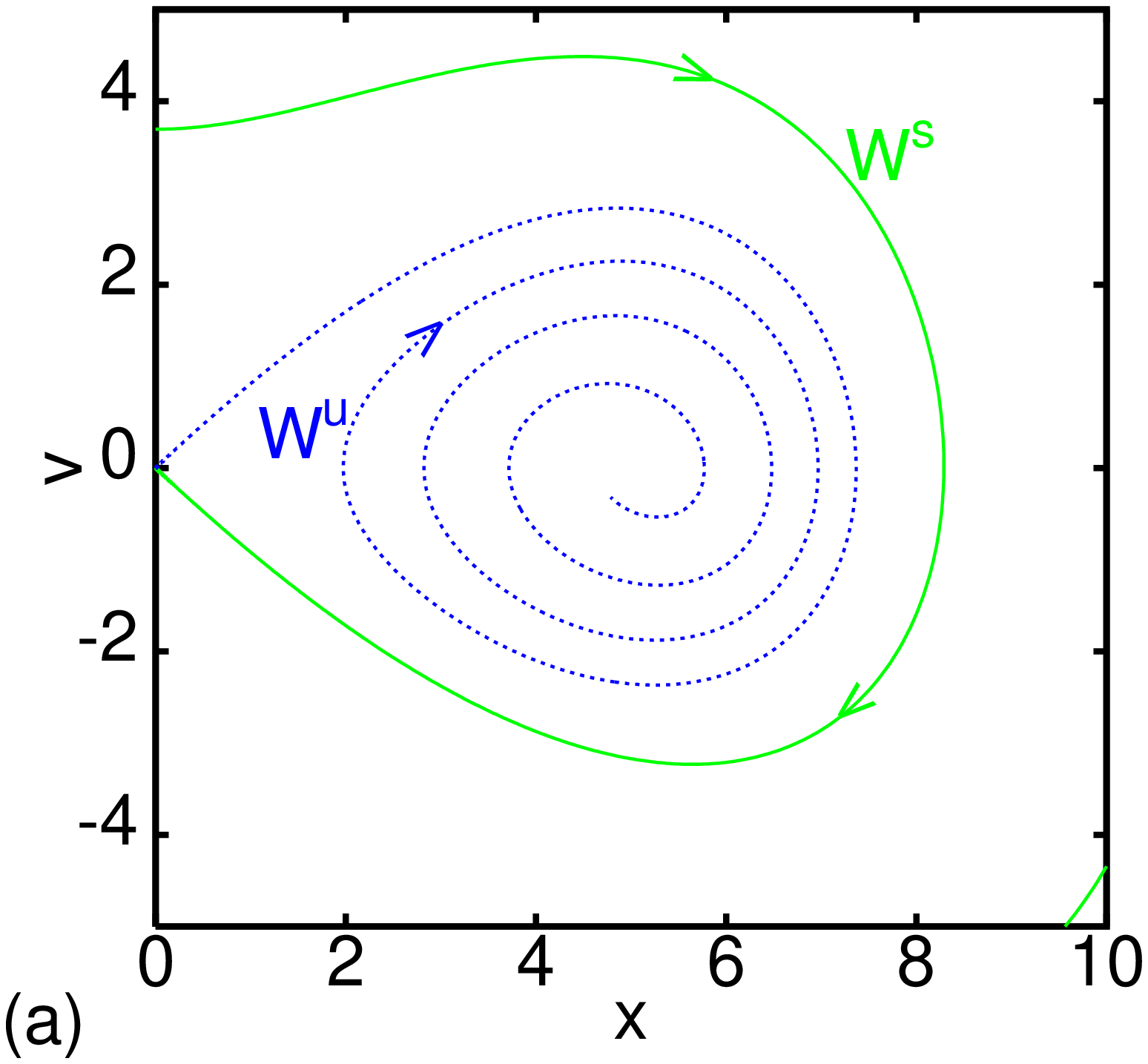,width=6.5cm,angle=-90} 
\epsfig{file=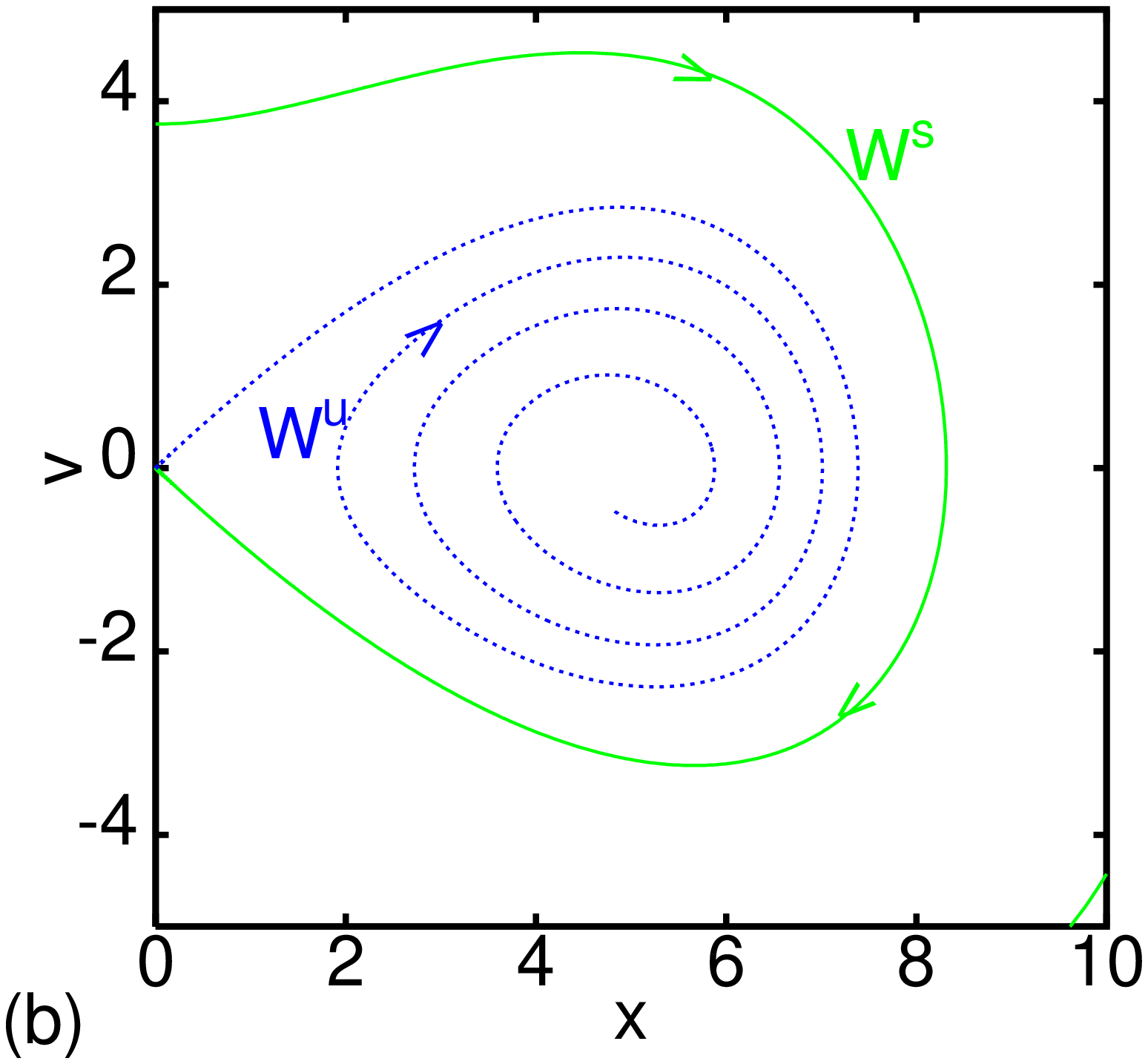,width=6.5cm,angle=-90}
}

\centerline{
\epsfig{file=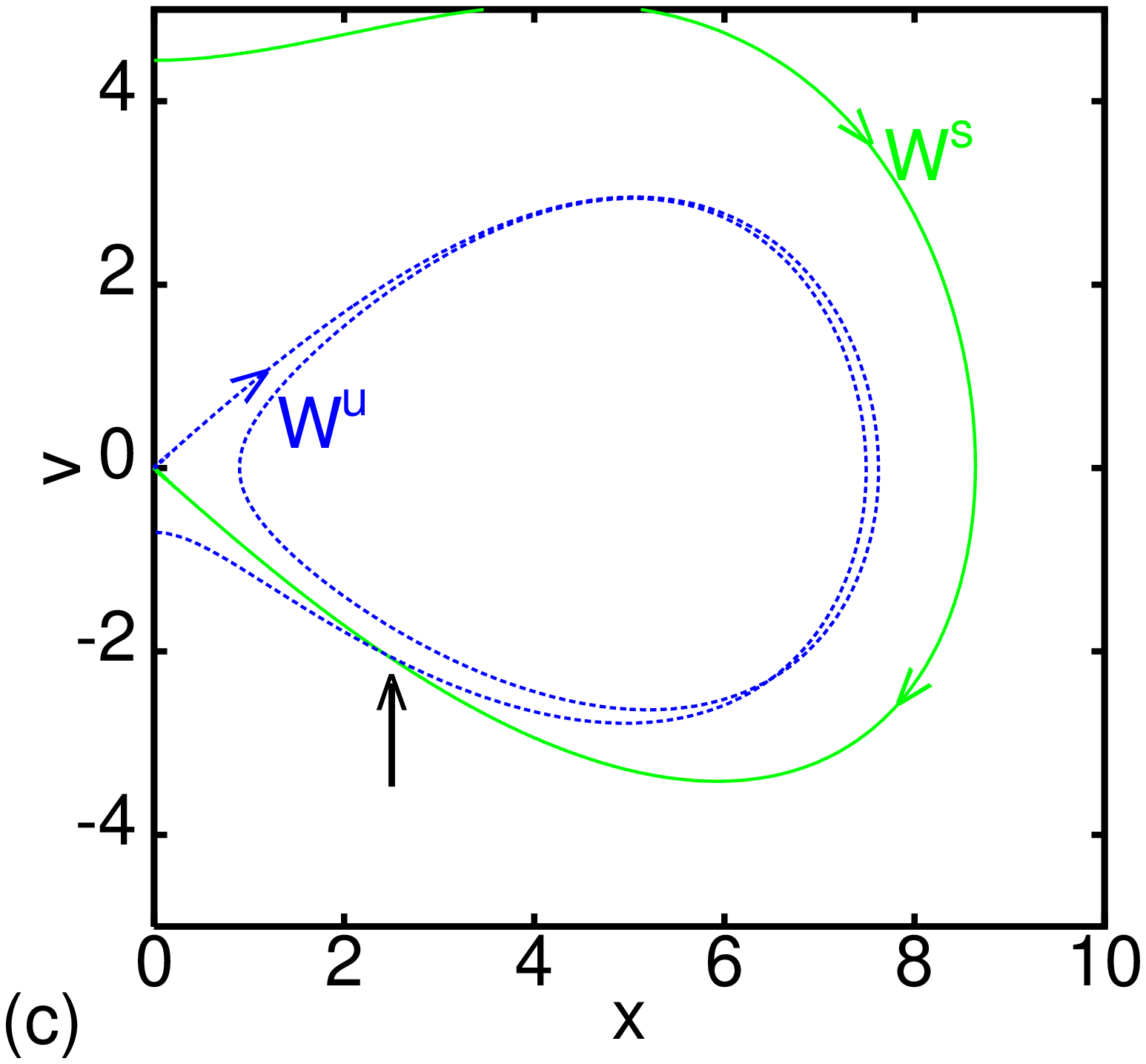,width=6.5cm,angle=-90}
\epsfig{file=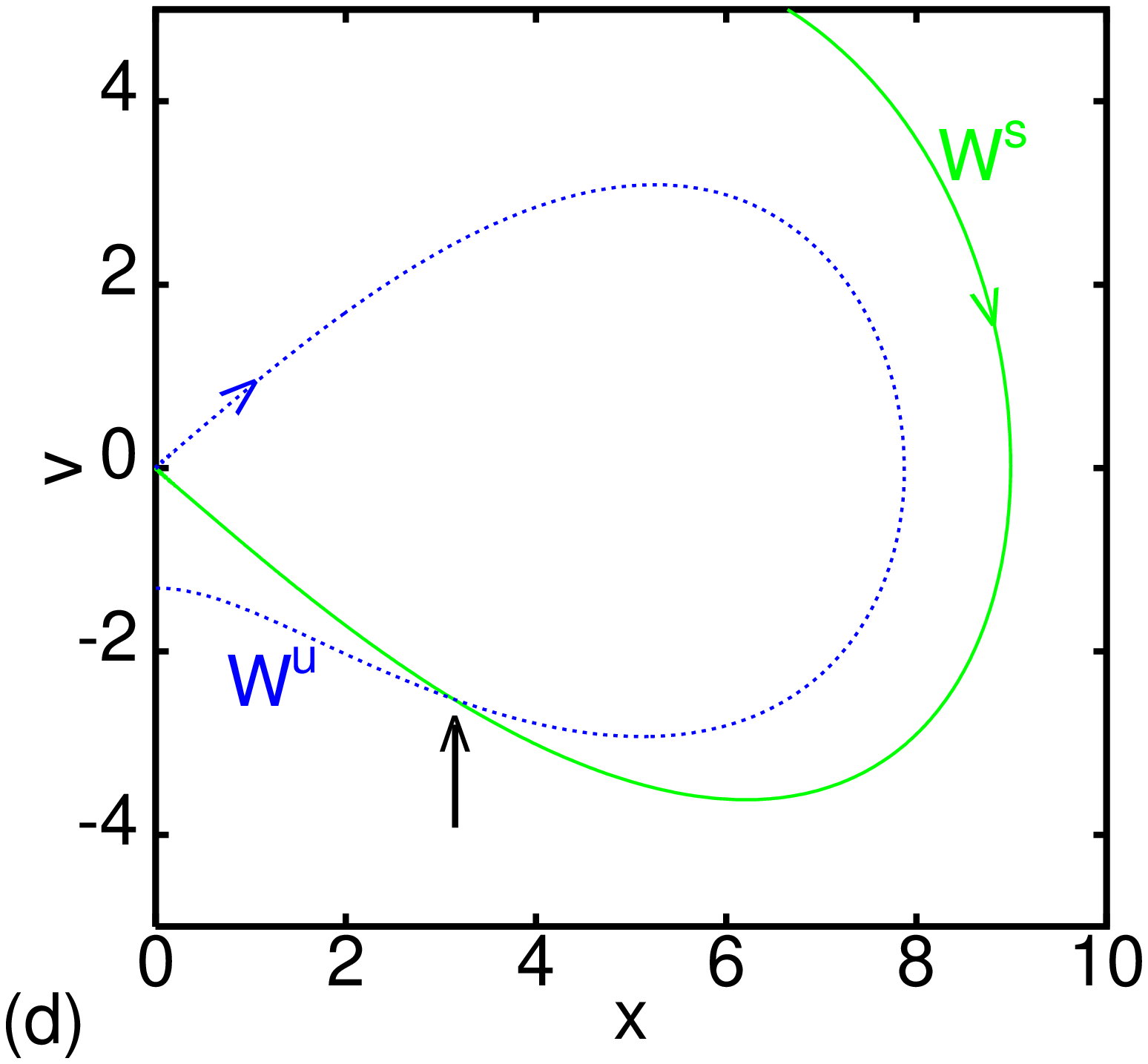,width=6.5cm,angle=-90}}

{\small \center Fig. 6. \label{fig6} 
Stable and unstable manifolds in presence of a perturbation
component ($\varepsilon=0.05$). Parameters used in calculations (Eq. 
\ref{eq2}):
$\omega=0.4$,
$\gamma=0.2$, $\delta=-1.0$, $\tilde \alpha=0.1$ and various excitation amplitude 
$\tilde \mu=0.87$ (Fig. 6a), 0.94 (Fig. 6b), 1.8 (Fig. 6c), 2.7 (Fig. 6d).
Note that $\mu= \varepsilon \tilde \mu$, $\alpha= \varepsilon \tilde 
\alpha$ (see Eq. \ref{eq3}).  
Arrows in Fig. 6c and d indicate the crossing points.}
\end{figure}

\begin{figure}[htb]
\centerline{
\epsfig{file=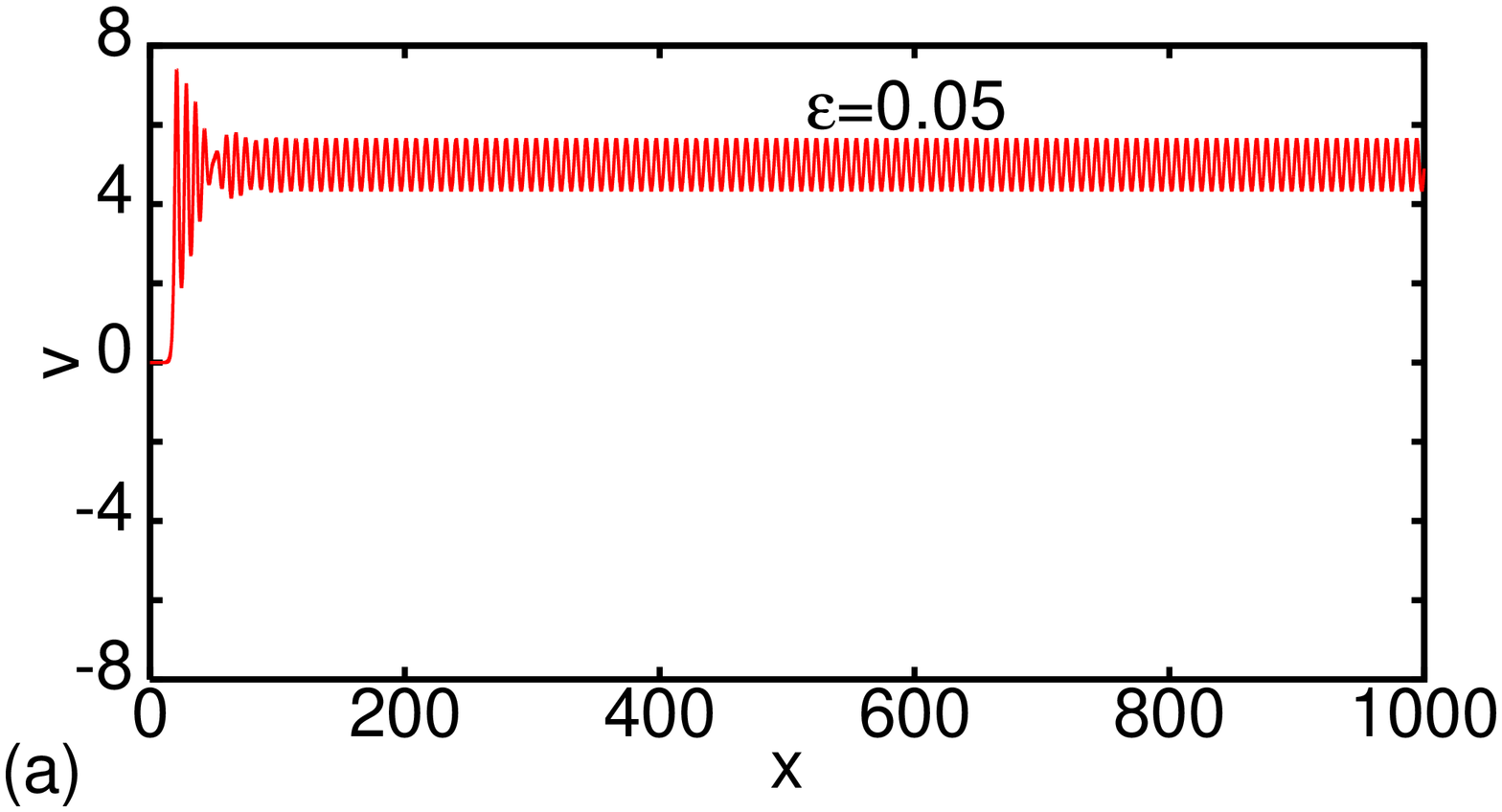,width=5.5cm,angle=-90}}

\centerline{
\epsfig{file=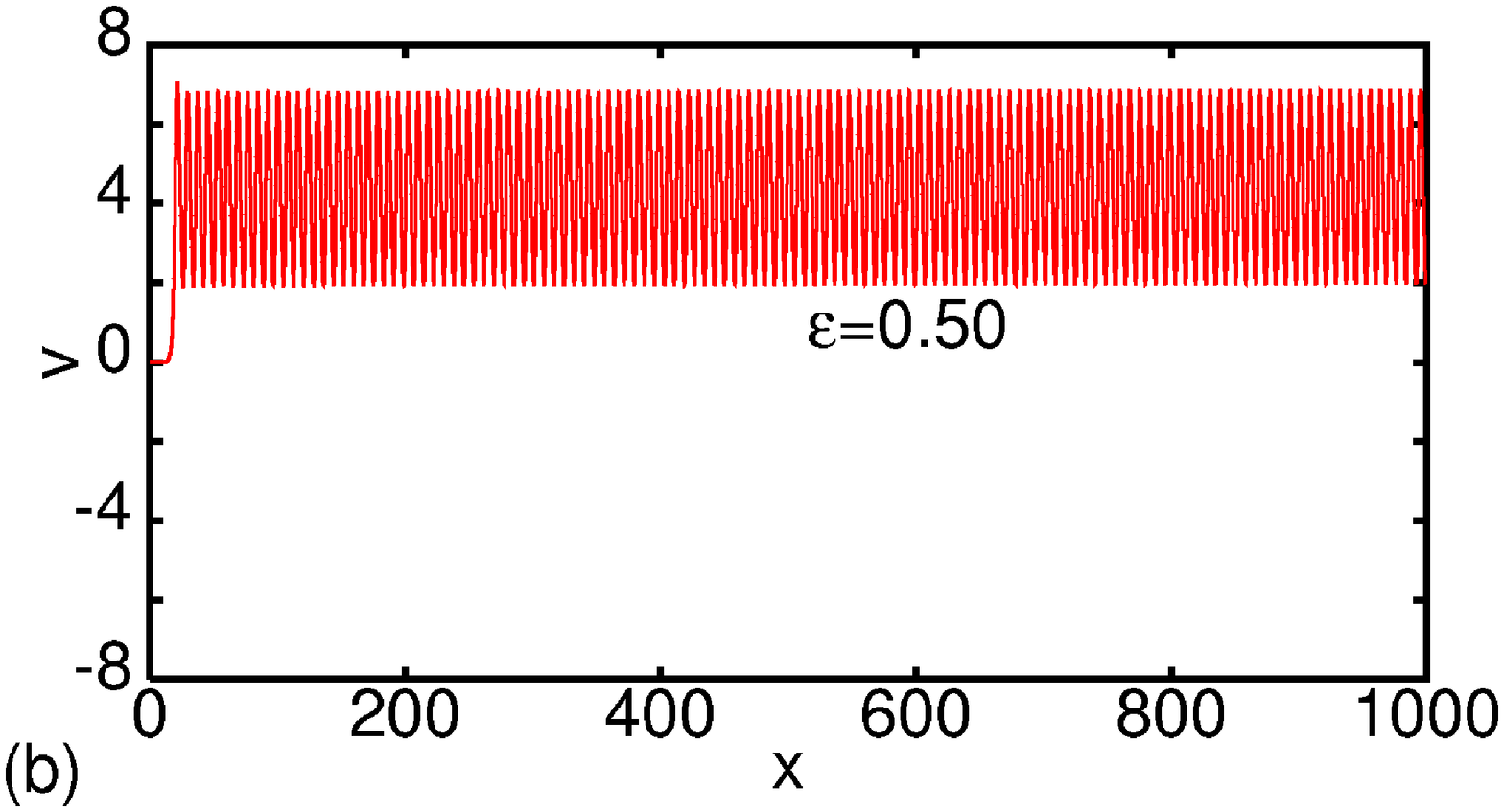,width=5.5cm,angle=-90}}

\centerline{
\epsfig{file=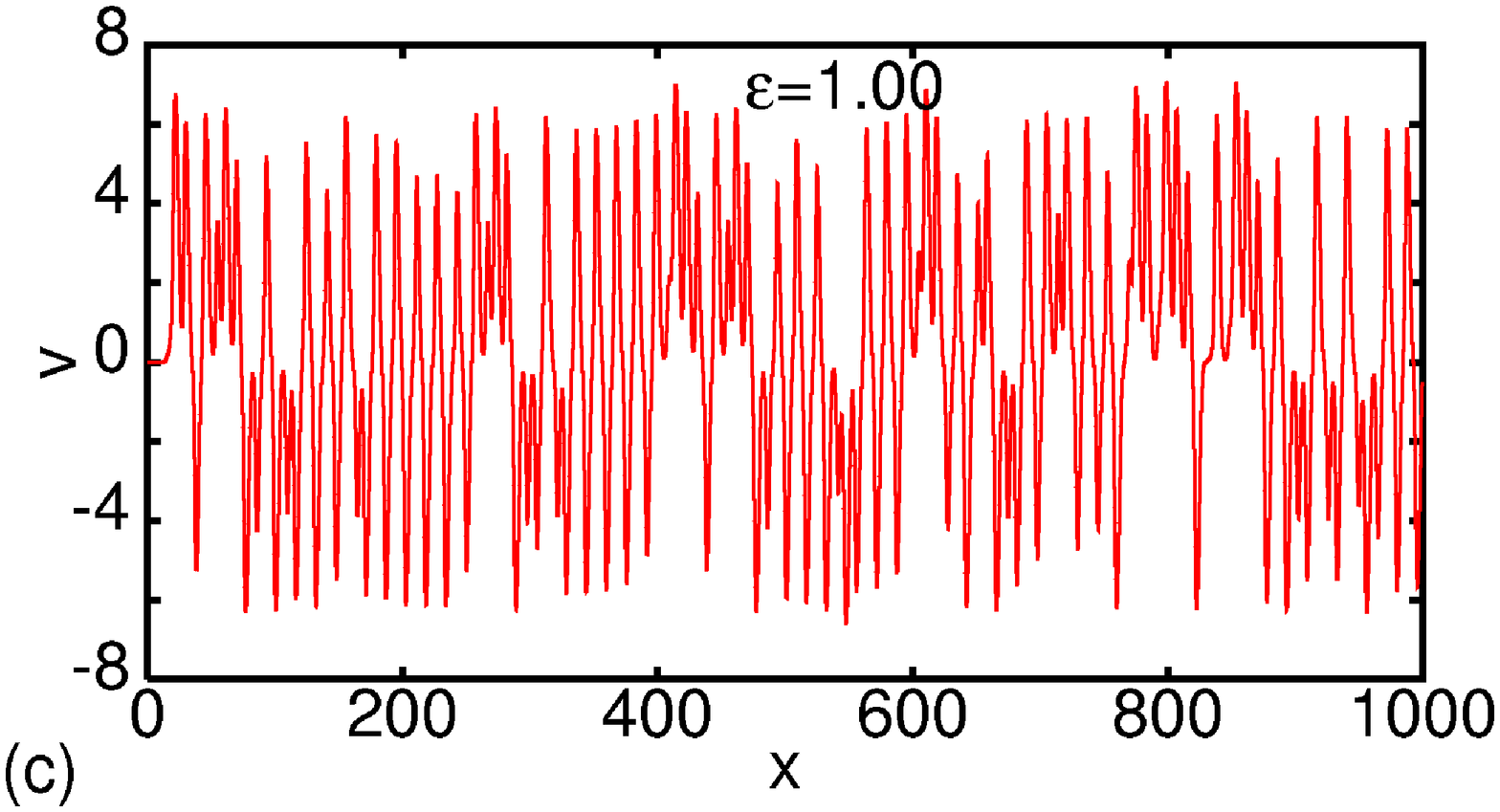,width=5.5cm,angle=-90}}

{\small \center Fig. 7. \label{fig7} 
Time histories $x(t)$ obtained by  simulation of Eq. \ref{eq2}
$\omega=0.4$,
$\gamma=0.2$, $\delta=-1.0$, $\tilde \alpha=0.1$,
$\tilde \mu=0.94$ and three values of $\varepsilon=0.05$ (Fig. 7a), $\varepsilon=0.50$ (Fig. 7b),
$\varepsilon=1.00$ (Fig. 7c). Note that $\mu= \varepsilon \tilde \mu$, 
$\alpha= \varepsilon \tilde
\alpha$ (see Eq. \ref{eq3}).
 
}
\end{figure}

\begin{figure}[htb]
\vspace{-0.7cm}
\centerline{
\epsfig{file=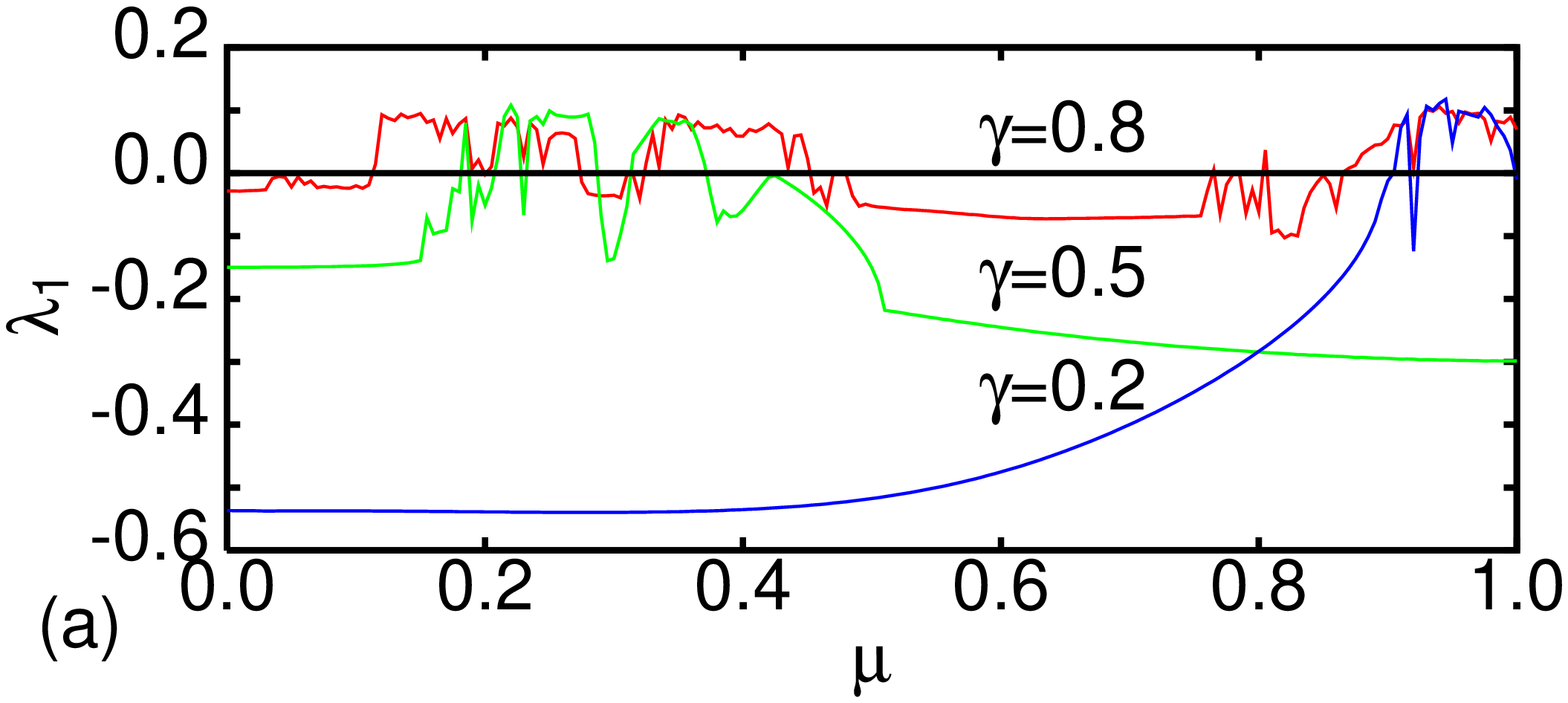,width=4.5cm,angle=-90}}

\vspace{-0.7cm}
\centerline{
\epsfig{file=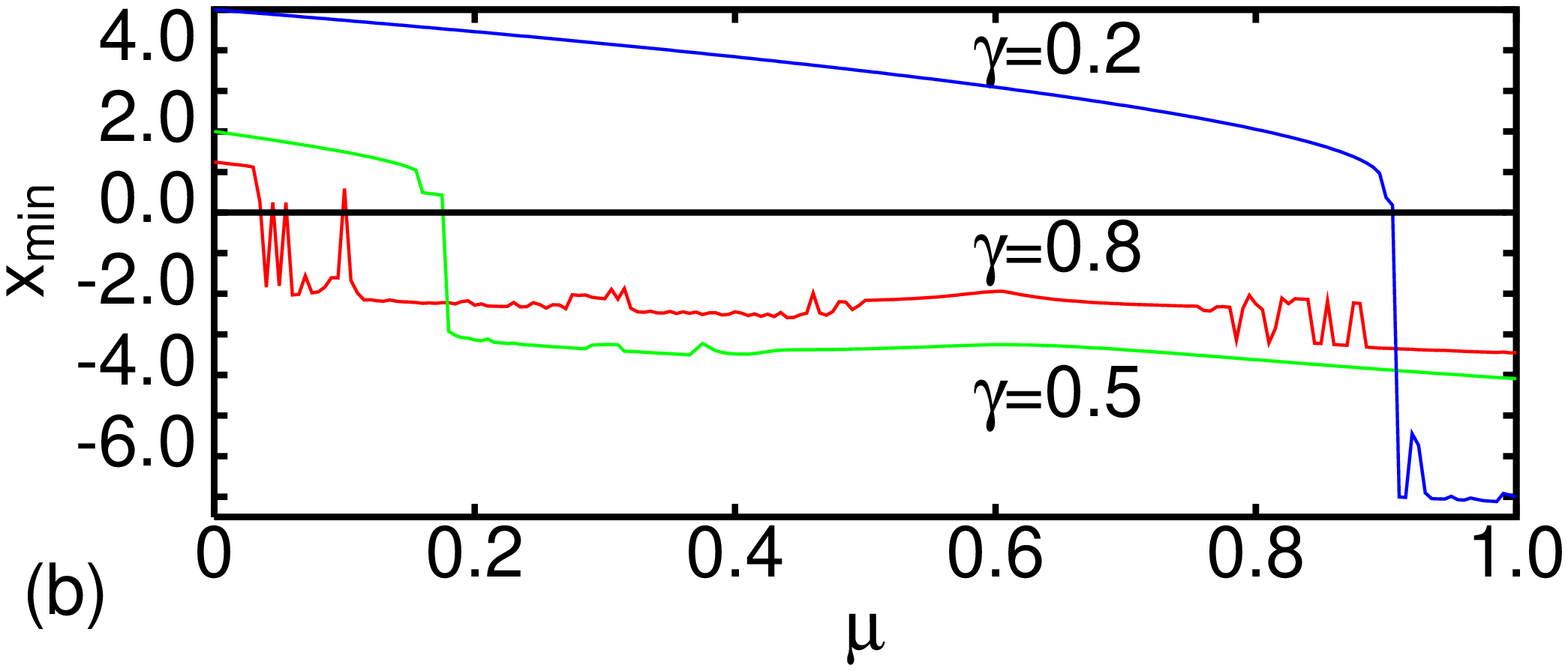,width=4.5cm,angle=-90}}

\vspace{-0.7cm}
\centerline{
\epsfig{file=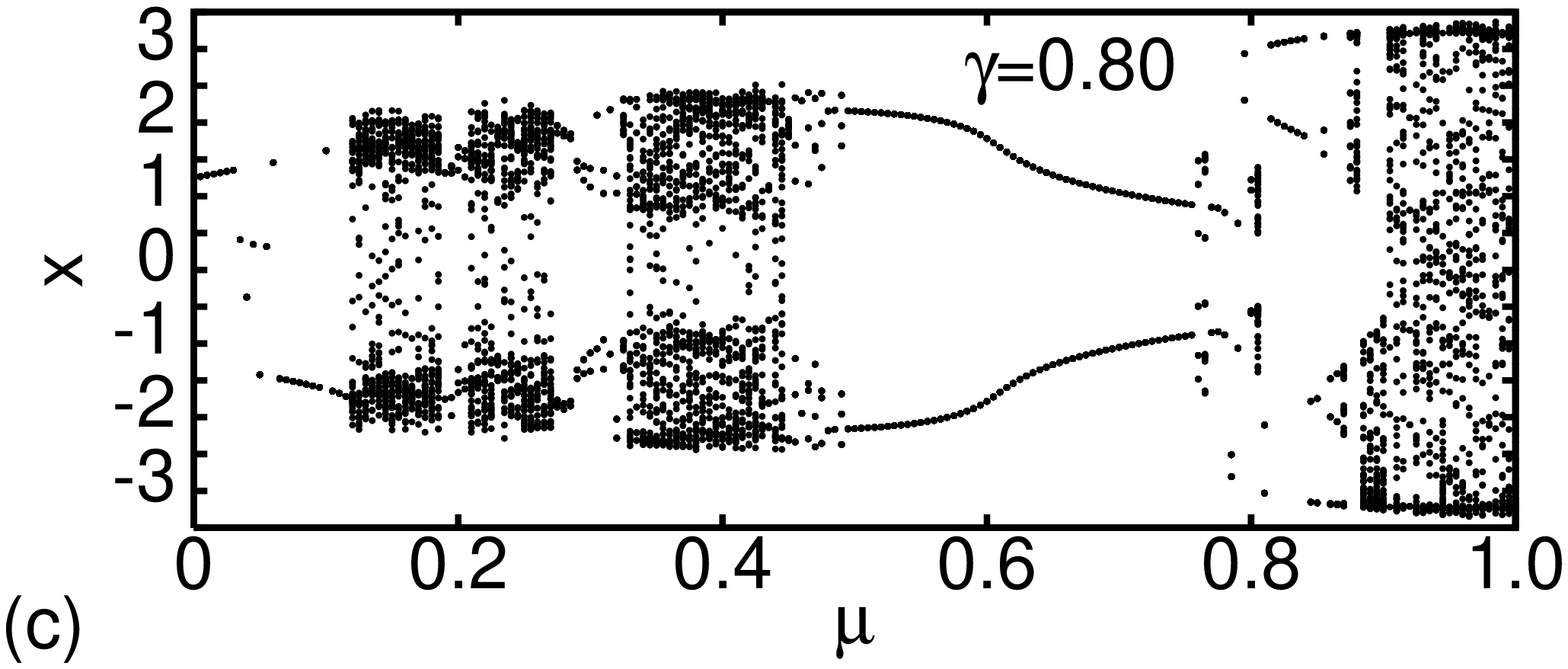,width=4.5cm,angle=-90}}

\vspace{-0.7cm}
\centerline{
\epsfig{file=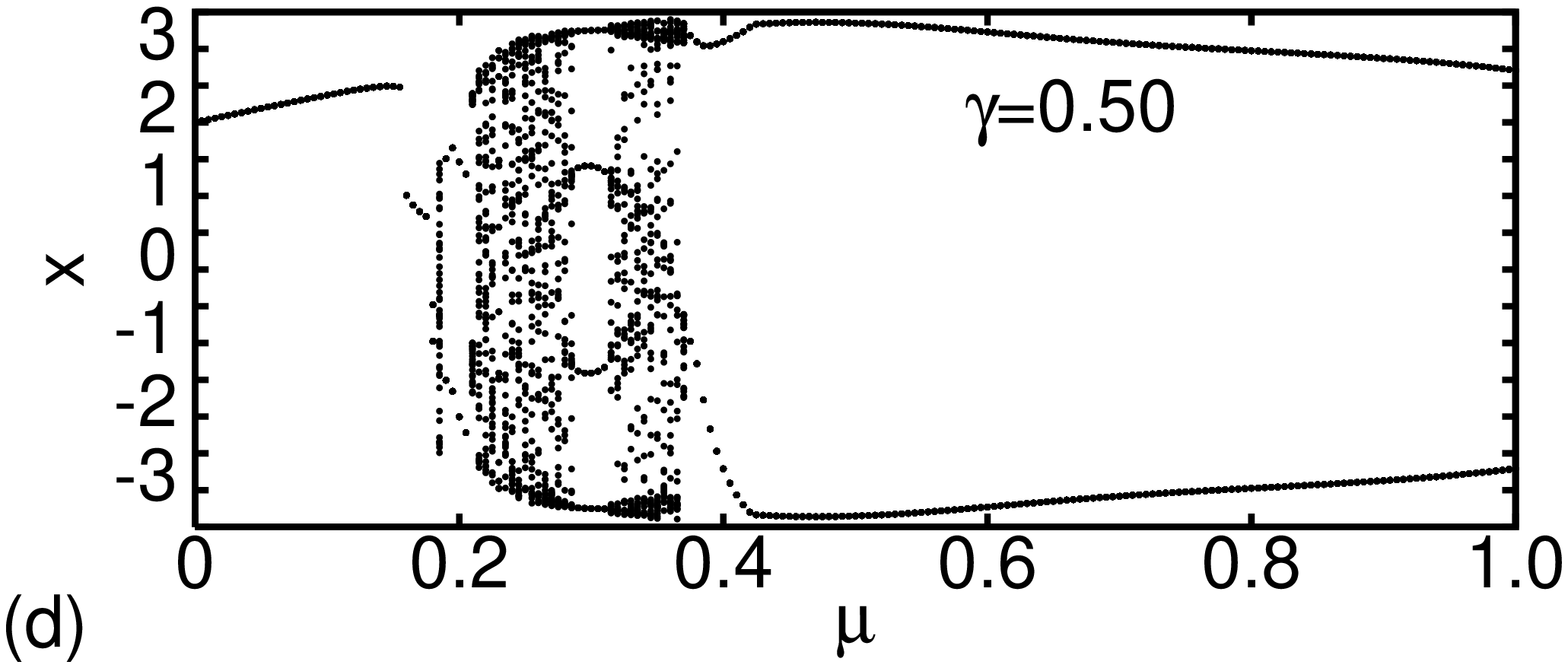,width=4.5cm,angle=-90}}

\vspace{-0.5cm}
\centerline{
\epsfig{file=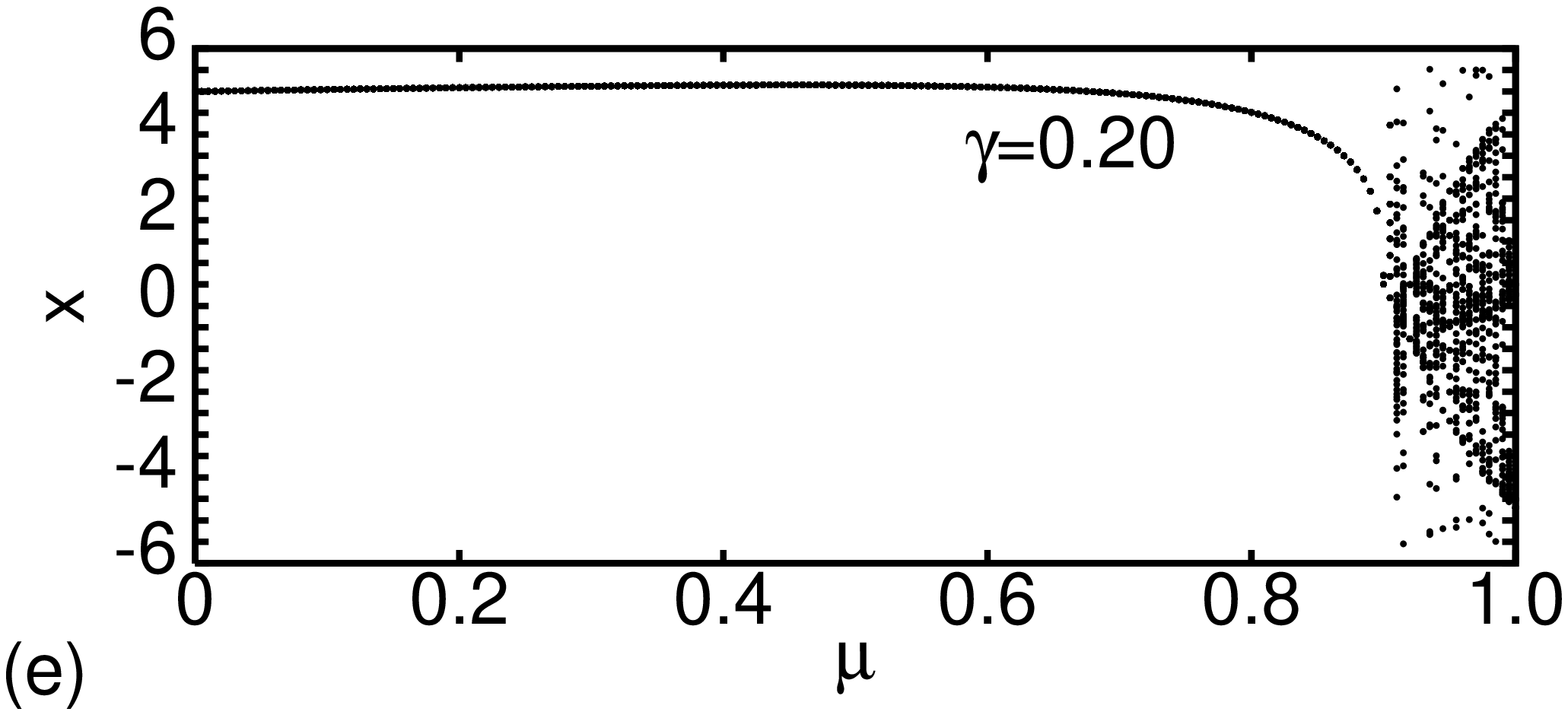,width=4.5cm,angle=-90}}

{\small \center Fig. 8. \label{fig8} 
Maximal Lyapunov exponent 
$\lambda_{1}$ (Fig. 8a) and the 
minimum 
point of 
displacement point $x_{min}$ (Fig. 8b) of steady state vibrations. 
Figs. 8d-e correspond to bifurcation 
diagrams for $\gamma=0.8$, 0.5 and 0.2 respectively for 
$\alpha=0.1$, $\delta=-1$ and $\omega=0.4$. Simulations were based on Eq.
\ref{eq1}. 
} 
\end{figure}

Naturally, from the above (Eqs. \ref{eq11}-\ref{eq13})
the Melnikov integral is given by:

\begin{eqnarray}
{\rm M}(t_0)  &=& \int_{- \infty}^{ + \infty} v^*(t-t_0) \left\{ \tilde 
\mu 
x^* (t-t_0) 
\cos{\left(2 
\omega 
t \right)} + \tilde \alpha \left[ 1- x^{*2}(t-t_0) \right]v^*(t-t_0) \right\} ~{\rm d} t \nonumber \\
&=& \int_{- \infty}^{ + \infty} v^*(t) \left\{ \tilde \mu 
x^* (t) 
\cos{\left(2
\omega
(t+t_0) \right)} + \tilde \alpha \left[ 1- x^{*2}(t) \right]v^*(t) \right\} ~{\rm d} t. 
\label{eq14}
\end{eqnarray} 

After substituting $x^*(t)$ and $v^*(t)$ given in Eq. \ref{eq10}  we obtain:

\begin{equation}
\label{eq15}
{\rm M}(t_0)= \tilde \mu I_{1}(t_0) + \tilde \alpha  I_{2},
\end{equation}
where the component integrals $I_1(t_0)$ and $I_2$ are defined: 

\begin{equation}
\label{eq16}
I_1(t_0) = \int_{-\infty}^{\infty}  v^*(t)x^*(t) \cos (2
\omega(t+t_0))   {\rm d} t
\end{equation}

and 

\begin{equation}
\label{eq17}
I_{2} = \int_{-\infty}^{\infty}  \left( 1- x^{*2}(t) 
\right)v^{*2}(t)  {\rm d } t.
\end{equation}

After evaluation of the above integrals (see Appendix B: Eqs. \ref{eqB.17} 
and \ref{eqB.4} ) we get 
the exact forms of $I_1(t_0)$, and $I_{2}$: 

\begin{eqnarray}
I_1(t_0) &=& \frac{12 \pi \delta^2}{\gamma^2} \frac{\omega^2 \left(1 + 4
\omega^2
\right)}{\sinh ( 2 \omega \pi)} \sin (2 \omega t_0) \label{eq18}
\\
I_{2} &=& \frac{\delta^2 \sqrt{-\delta}}{\gamma^2} \left(
-\frac{6}{5}+\frac{72 \delta^2}{70 \gamma^2}\right). \label{eq19}
\end{eqnarray}

Surprisingly 
the other chose of system variables ($x$, $w$) lead to the same form of 
Melnikov function (Appendix A), Eqs. \ref{eq14}-\ref{eq17}.  
That means that both  perturbation procedures give the same result 
in the first approximation. This is partially due to the fact that in both 
cases we started from the same set of equations of unperturbed 
Hamiltonian $H^0$ (Eqs.  \ref{eq2},\ref{eq6}). 

Finally, the condition for a  transition to chaotic motion as a global homoclinic transition 
corresponding to a 
horse-shoe 
type of stable and unstable manifolds 
cross-section, can be written as

\begin{equation}
\label{eq20}
{\displaystyle \bigvee_{t_0}} 
~~~ {\rm M}(t_0)=0 {\rm ~~ and ~~} 
\frac{\partial {\rm M}(t_0)}{\partial t_0} \neq 0. 
\end{equation}
The above integrals (Eq. \ref{eq18},\ref{eq19})
 together with the last condition (Eq. \ref{eq20}) yields   
a critical value of excitation amplitude 
$\mu_c$:

\begin{equation}
\label{eq21}
 \mu_c= \frac{\alpha \sqrt{-\delta} \sinh ( 2 \omega \pi)}{12 \pi \omega^2 
(1+ 4 \omega^2)} 
\left|
-\frac{6}{5}+\frac{72 \delta^2}{70 \gamma^2}\right|. 
\end{equation} 

for which stable and unstable manifolds cross. 

Namely we get the critical amplitude $\mu_c$ versus frequency 
$\omega$, and parameter $\gamma$ 
which is plotted in Figs. 4a--d. Above this value $\mu > \mu_c$
the system transit through a global homoclinic bifurcation 
which is a necessary condition  for appearance of chaotic vibrations. 
Interestingly, larger $\gamma$ leads 
to smaller $\mu_c$ (Fig. 4d) Note also, that the condition 
(Eq. \ref{eq21}) 
is based on results of
 perturbation procedure in its lowest order approximation.
Equation \ref{eq21} is the main results of our investigation.
\\ \\

\section{ Numerical simulations}

To verify the analytical findings we have performed series of
numerical simulations
of Eq. \ref{eq1}.
In Fig. 5 we show phase diagrams and simultaneously  corresponding
Poincare
sections stroboscopic points collected with frequency $\Omega=2\omega$ for
chosen sets initial system (Eq. \ref{eq1})
parameters: $\omega=0.4$, $\gamma=0.8$ $\mu=0.02$, 0.10 (Fig.
5a) and
$\mu=0.116$ (Fig. 5b); $\omega=0.4$, $\gamma=0.5$ and $\mu=0.160$ (Fig.
5c) and $\mu=0.185$
 $\omega=0.4$, $\gamma=0.2$ (Fig. 5d); $\mu=0.87$ (Fig.
5e); and $\mu=0.94$ Fig. 5f); $\omega=0.45$, $\gamma=0.2$ and
$\mu=0.90$ (Fig.
5g); and $\mu=0.97$ Fig. 5h).
The structure of the examined attractors as well as calculated Lyapunov
exponents enables to classify the dynamics of the system
(see Fig. 5 and figure
caption).
For comparison with analytical results we
indicated the simulated cases by points, showing the types of vibrations:
R (regular) and C (chaotic). 
For the last case for relatively small $\gamma$ ($\gamma=0.2$ in Fig. 4c, 
and Figs. 5 
e--h) one can see some 
discrepancy between simulated data and analytical
results.
Analytical results indicate that the homoclinic transition take place  
for $\mu_c > 1$ but simultaneously for the same system parameters numerical 
results states that $\mu_c < 1$ is enough to transit into chaotic 
vibrations ($\mu_c \approx 1.5$, Fig. 4c). This
could be effect of relatively large $\mu$ in perturbation procedure (Eq. \ref{eq3}).
In the above calculations we used the same
initial conditions  $(x_{in},v_{in})=(0.45,0.1)$. 

The interesting discrepancy between analytical (Fig. 4c) and numerical simulation (Fig. 5e-f) results
deserved some additional comments. In this aim we plotted, in Fig. 6)  our results on simulations of 
perturbed system (Eq. 
\ref{eq3}) related to stable and unstable orbits. Because the Melnikov 
theory formulated in the lowest order approximation 
is valid in the limit of small $\varepsilon$  ($\varepsilon \rightarrow 0$) 
we decided to use a relatively small value of this parameter $\varepsilon=0.05$. 
Other parameters used in calculations (Eq. \ref{eq3}) are as follows
$\omega=0.4$,
$\gamma=0.2$, $\delta=-1.0$, $\tilde \alpha=0.1$ and various excitation amplitude
$\tilde \mu=0.87$ (Fig. 6a), 0.94 (Fig. 6b), 1.8 (Fig. 6c), 2.7 (Fig. 6d).
Note, in cases of Fig. 6a and b, which correspond according to Figs. 4c, 5e, 5f, to
regular and chaotic motions for $\varepsilon=1$ (original equation Eq. 
\ref{eq1}). Obviously there are no crossing
points between the stable $W^S$ and unstable $W^U$ manifolds. 
After comparing these results
it is obvious that in this particular case the final result of our consideration 
could depend on $\varepsilon$. In fact the difference appears between the limit $\varepsilon \rightarrow 0$ (or  
$\varepsilon=0.05$ in Figs. 6a,b)
and  $\varepsilon=1.0$, what is not surprise taking into account relatively large value of 
$\tilde \mu$ ($\tilde \mu 
\approx 1.0$  Fig. 4c). Interestingly for the present value of  $\varepsilon$ ( $\varepsilon=0.05$)
our simulations (Fig. 6c and d) confirm analytical predictions (see the solid curve in Fig. 4c). As $\mu > 
\mu_c \approx 1.5$ in both cases we observe crossing points of the stable 
and unstable orbits (indicated by arrows in 
Fig. c and d). Evidently increasing $\tilde \mu$ causes faster meeting of corresponding curves $W^S$ 
and 
$W^U$, as one can see Fig. 6c and Fig. 6d differ by winding numbers $N$ of motion necessary for such 
crossing:  
$N= 1$ for  Fig. 6c  and $N= 2$ for  Fig. 6d.  
We have also done additional tests for other system 
parameters including different $\gamma=0.5$, 0.8 (as in Fig. 4a-b and 
Figs. 5a-d) and  $\omega=0.45$ (as in Fig. 4c and 
Figs. 5g-h). In all cases we have also got agreement 
with the corresponding 
analytical curves (Fig. 4).    
   In the next figures (Fig. 7a-c) we show the evolution of the oscillation with increasing
$\varepsilon$. Here we have plotted time histories for different values of $\varepsilon$. 
Starting from  vicinity of the saddle point  we obtained  regular history (Fig. 7a) for $\varepsilon=0.05$.
For larger value of control parameter $\varepsilon$ ($\varepsilon=0.5$ in Fig. 7b) 
we observe enlargement of the 
amplitude of these motion and finally its transition to chaotic motion for a large enough $\varepsilon$
($\varepsilon=1.0$ in Fig. 7c).

For better clarity we decided to show the transition to chaos through
 the maximal Lyapunov exponent $\lambda_1$ and bifurcation diagrams for 
three values of 
$\gamma$ ($\gamma=0.2$, 0.5, 0.8) versus $\mu$ (Fig. 8).
Positive value of $\lambda_1$ in Fig. 8a( see the region $\sim 0.15 < \mu 
< \sim 0.45$ for $\gamma =0.8$) as well as black regions in 
corresponding bifurcation 
diagrams (Fig. 8c-e) imply transition to chaos. This is a way of numerical identification 
of critical value $\mu_c$ regarding a local bifurcation. In case of 
$\gamma=0.8$ and 0.5 we observe the confirmation of previous findings
for numerical (Poincare maps Figs. 5a-d) and analytical results
(Fig. 4a-b) while the case of $\gamma=0.2$ is different.
Here again we observe  some disagreement between numerical (Figs. 
5e-h) and 
analytical results (Fig. 4c). Lyapunov exponent method gives smaller 
value of $\mu_c$.
 The results, obtained while simulating Eq. \ref{eq1}, show the actual 
critical 
values of $\mu$ for
transition to chaotic vibrations. 
Note also, that in many cases the observe the 
chaotic oscillations existing in both potential wells (with positive and 
negative $x$). However this behaviour is not a rule because Figs. 5g,d show 
the transition to chaotic behaviour inside the single well $x> 0$.
To explore this effect more systematically we plot (in Fig. 8b) 
$x_{min}$ versus $\mu$ and three values of 
$\gamma=0.2$, 0.5, 0.8.   It appeared that for $\gamma=0.8$ escapes 
and 
returns of the considered system
to a single well oscillatory motion can happen for an interval 
of $\mu$ 
parameter which is close but slightly below the value transition to 
chaotic vibration. This means that the system has already possessed double 
well attractor just before its transition to chaotic behaviour.    
Note, in our numerical calculations, to have some direct insight into the 
system capability of transition to chaotic vibration and its escape from 
the a single well and to minimize the time of our calculations, 
the initial conditions at the beginning 
($\mu=0$) were assumed to be  $(x_{in},v_{in})=(0.45,0.1)$ and 
continuing the calculations we used 
final displacement and velocity obtained for the former value 
of $\mu$  $(x_{fi},v_{fi})$ as initial 
conditions for any new value of $\mu$. 
We also note, that this system resembles the other examined by 
Thompson \cite{Thompson1989} with the same type of nonlinearity of square 
type but without Van 
der Pol term and external forcing instead of our parametric one. In his 
study he has  got the condition for a global 
homoclinic  transition just before the escape from the single potential 
well. In our case the effect of self excitation, connected with the Van 
der Pol damping term 
is stronger for $\gamma=0.8$ 
(Fig. 1).
This effect can be deduced from the excitation term itself for  
$x \rightarrow 0$ we have ${\rm VdP}(x,\dot{x}) \rightarrow -\alpha \dot{x}$ which means
negative damping if the system moves close to $x=0$. The extra energy 
generated by such a 
negative 
damping makes it possible to overcome the potential barrier between two 
symmetric potential wells much easier comparing the case without 
self excitation  effect.
On the other hand,  the size of 
attractor for given $\gamma$ (Fig. 5) is solely determined by the shape of 
the  nonlinear potential
(Eq. \ref{eq2}, Fig. 1). For $\gamma=0.8$ we have also found that for $\mu > \mu_c$
a transition chaotic vibrations was preceded by transient chaotic behaviour.
The steady state motion undergo  escape from a potential well with formation of 
symmetric attractor spreading on two potential wells. With increasing $\mu$ the symmetry of
attractor was broken and the system transit a series of period doubling bifurcations 
as was discussed in  earlier papers \cite{Szemplinska1993,Szemplinska1995,Tyrkiel2005}.

\section{Summary and Conclusions}
In summary we have studied conditions of a 
global homoclinic bifurcation in a double well potential --van der Pol 
system with parametric excitation.
Such a bifurcation  correspond to transit chaotic behaviour of the 
system and with some further increasing of the excitation amplitude can 
lead to the permanent chaos \cite{Szemplinska1993,Szemplinska1995,Kapitaniak1991,Tyrkiel2005}.

Using the Melnikov method we have got the analytical formula for
transition to chaos in a one degree of freedom, system subjected
to parametric excitation
with a non-symmetric
stiffness with self-excitation term. 
In our case this effect
is mutually introduced through the Van der Pol damping
and  parametric excitation terms. Our analytical results 
are consistent with direct 
computations on homoclinic orbits.  

Note that our vector field (Eq. \ref{eq3}) is of $C^1$ class due to 
non-continuity of the 
second derivative at $x=0$ line  or a piece--wise
$C^2$ smooth system.
In fact he standard theory \cite{Guckenheimer1983} 
assumes that expansion in Taylor
series in $\varepsilon$ is good to second order at least (for uniform 
bounds 
on $\varepsilon^2$ term). This requires
that
$h(x,\dot x)$
and $g(x,\dot x)$ 
are of $C^2$ class with respect to $x$ and $\dot x$,
and  the Hamiltonian is of $C^3$ class. However for 
specific 
cases the theory can be applied for
weaker assumptions.
For any non-smooth systems one should  check that piece--wise smooth 
solutions
can be assembled and indicate  any jumps in derivatives that can occur 
\cite{Holmes2005}.

Let us write the forms $h(x,\dot x)$ and $g(x,\dot x)$ (Eqs. 
\ref{eq12} 
and \ref{eq13}) as vectors:
\begin{eqnarray}
{\bf h} &=& \left[\left(\delta x + \gamma |x^*|x^*\right),  v 
\right]=[h_1,h_2], \nonumber \\ 
{\bf g}&=&
\left[\left( \tilde{ \mu} x \cos{2 \omega \tau} + \tilde{\alpha}
\left(1-x^2 \right)
v \right),0 \right]. \label{eq22}
\end{eqnarray}

Now, the Mielnikov function M($t_0$) can be treated as a 
projection of the vector 
${\bf h}^{\perp}=[-h_2,h_1]$ into the
${\bf g}$ direction (a scalar product), Namely
\begin{equation}
\label{eq23}
{\rm M}(t_0)=\int_{-\infty}^{\infty}  {\bf g} (t)\cdot {\bf 
h}^{\perp}(t+t_0)~{\rm d} t.
\end{equation}
Note that  the vector element $h^{\perp}_2$ ($h^{\perp}_2=h_1$) is 
projected out the Melnikov integral.
Moreover the same argument applies to any $x$ derivative of $h_1$ making 
its 
non-smooth behaviour at $x=0$
unimportant for the Melnikov theory application. In other words this 
non-smoothness at $x=0$ is not likely to produce
any jump to the homoclinic orbit (Fig. 3).

Interestingly the families of function $\mu_c (\omega)$ plotted 
against $\omega$ scales as $\sim \gamma^{-2}$ leading to small $\mu_c$ 
for 
relatively large $\gamma$ ($\gamma=0.8$) and much larger $\mu_c$ for 
$\gamma=0.2$. To confirm these results we have performed numerical
simulations showing corresponding phase diagrams, Poincare maps 
Lyapunov exponents and bifurcation maps.  The Lyapunov exponent has been
calculated using the algorithm provided by Wolf {\em et al.} 
\cite{Wolf1985}.

We have  noticed some discrepancy (due to relatively large values of $\mu_c$: $\left|\mu_c/\delta 
\right| > 1$),
between simulated data and 
analytical 
results, especially,
in case of small $\gamma$. 
It seams that our Melnikov analysis provides
results which are correct up to the first order. The results
can be possibly improved  in higher order
approximations \cite{Lenci2005}
The other possibility is  the stronger influence of 
the Van der Pol term on the system dynamics. However,
keeping the Van der
Pol term unchanged, 
we have not analyzed this effect  so deep as it deserves
leaving this aspect to future studies.

\section*{Acknowledgements}
This paper has been partially  supported by the  Polish Ministry of
Science and Informatization.
GL would like to thank the Max Planck
Institute for the Physics of Complex Systems for hospitality.
Authors would like to thank Prof. P. Holmes for discussion and unknown 
reviewers for valuable comments.

\appendix{\Large \bf \noindent Appendix A}
\def\thesection{A}
\setcounter{equation}{0}
\def\theequation{A.\arabic{equation}}  

The starting equation of motion (Eq. \ref{eq1}) is given as a differential equation of the first order. To 
perform further analysis of the system it must be split
 into two equations of the second order. The usual way based on time derivatives of displacement and 
velocity ($x,v$) was discussed in Sec. 2 (Eq. \ref{eq2}).
The other   
concept (Eq. \ref{eq3}), connected to division on 'slow' and 'fast' variables  leads to a 
another pair of equations:  
\begin{eqnarray}
\label{eqA.1}
& & \dot{x} = w -  \varepsilon\tilde{\alpha} \left (
x-\frac{x^3}{3}  \right) \\
& & \dot{w}  = -\delta x - \gamma |x|x + \varepsilon  \tilde{\mu} \left(
 \cos{\left(2 \omega t \right)}
\right). \nonumber
\end{eqnarray}
The above splitting does not effect 
the unperturbed Hamiltonian $H^0$ which is of the same form as for (Eq. \ref{eq2}) Thus $w$ has the same 
meaning of $v$ but perturbations are now 
appearing in both equations instead of one.

Now, the gradient of unperturbed Hamiltonian:
\begin{equation}
\label{eqA.2}
h = \left(\delta x + \gamma |x^*|x^*\right) {\rm d} x  + v {\rm d}v,
\end{equation}
while  $g'$ is a perturbation form

\begin{equation}
\label{eqA.3}
g' = \left( \tilde{ \mu} x \cos{2 \omega \tau} +  \right) {\rm d}x -
\tilde{\alpha} \left (
x-\frac{x^3}{3}  \right) {\rm d}v.
\end{equation}
defined on corresponding stable or unstable 
manifolds $(x,v)=(x^*,v^*)$ (Eq. \ref{eq10}, Fig. 
2).
The Melnikov function:
\begin{eqnarray}
\label{eqA.4}
M'(t_0) &=&
\int_{- \infty}^{ + \infty}  h( x^*, v^*)  \wedge g'( x^*,
v^*) {\rm d} t \\
&=& \int_{- \infty}^{ + \infty} 
\left(
\tilde \mu
x^* v^*
\cos{\left(2
\omega
(t+t_0) \right)} + \tilde \alpha \left(\delta x^* + \gamma x^{*2}\right)
\left (
x^*-\frac{x^{*3}}{3}  \right)
\right) {\rm d} t \nonumber
\end{eqnarray}
 can be evaluated after  substituting $x^*(t)$ and $v^*(t)$ by formulae 
given in Eq. \ref{eq10}:
\begin{equation}
\label{eqA.5}
M'(t_0)= \tilde \mu I'_{1}(t_0) + \tilde \alpha  I'_{2},
\end{equation}
where
\begin{equation}
\label{eqA.6}
I'_1(t_0) = \int_{-\infty}^{\infty} {\rm d} t~ v^*(t)x^*(t) \cos (2
\omega(t+t_0))
\end{equation}
and
\begin{equation}
\label{eqA.7}
I'_{2} =\int_{-\infty}^{\infty} {\rm d } t  \left(\delta x^*(t) + \gamma
x^{*2}(t)\right)
\left (
x^*(t)-\frac{x^{*3}(t)}{3}  \right).
\end{equation}

Note the expression for $I'_1(t_0)$ (Eq. \ref{eqA.6}) coincides exactly with 
 $I_1(t_0)$ analyzed in Sec. 2 (Eq. \ref{eq16}) while $I'_2$ can be transformed
exactly to $I_2$ if one make use of the definition of the homoclinic orbit
parametrisation (Eqs. \ref{eq7}-\ref{eq8}) and substituting: 
\begin{equation}
\label{eqA.8}
\frac{{\rm d}v}{{\rm d}t}=-\delta x - \gamma |x|x
\end{equation}
 into the expression Eq. \ref{eqA.7} and integrating it by parts: 
\begin{eqnarray}
I'_{2} &=& -\int_{-\infty}^{\infty} {\rm d } t \frac{{\rm d}v}{{\rm 
d}t} \left (
x^*(t)-\frac{x^{*3}(t)}{3}  \right), \\ &=& \int_{-\infty}^{\infty} 
{\rm d } t \left( 1- x^{*2}(t)
\right)v^{*2}(t)=I_{2}.  
   \end{eqnarray}
Thus the above  Melnikov function form $M'(t_0)$ coincides exactly the 
the form of ${\rm M}(t_0)$ obtained Sec. 2 (Eqs. \ref{eq14}--\ref{eq19}).

\appendix{\Large \bf \noindent Appendix B}
\def\thesection{B}
\setcounter{equation}{0}
\def\theequation{B.\arabic{equation}}  

\begin{figure}[htb]
\centerline{
\epsfig{file=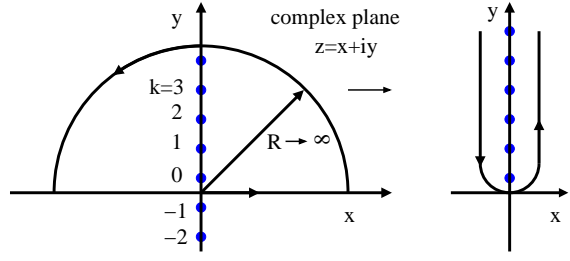,width=7.5cm,angle=0}}
\caption{
Deformed contour integration schema and imaginary poles.   
}
\end{figure}

Evaluation of the integral $I_2$ is straightforward. After substitution 
$x^*(t)$, and $v^*(t)$ (Eq. \ref{eq10}) we have:

\begin{eqnarray}
\label{eqB.1}
I_2 &=&\int_{-\infty}^{\infty}
 \left( 1- x^{*2}(t)
\right)v^{*2}(t) {\rm d } t \\ &=&
\frac{9}{4}\frac{\delta^3}{\gamma^2} \int_{-\infty}^{\infty}
\frac{\tanh^2\left(\frac{\sqrt{-\delta}t}{2}\right)}{\cosh^4
\left(\frac{\sqrt{-\delta}t}{2}\right)} \left( 1-
\frac{9}{4}\frac{\delta^2}{\gamma^2} \left(
\tanh^2\left(\frac{\sqrt{-\delta}t}{2}\right) -1 \right)^2\right) {\rm d } t.
\nonumber \end{eqnarray}

and simple algebraic manipulations:
\begin{equation}
\label{eqB.2}
t=\frac{2 \tau}{\sqrt{-\delta}}, ~~~~~~~ 
\tanh \tau=\xi
\end{equation}
we obtain
\begin{equation}
\label{eqB.3}
I_2=-\frac{9}{2}\frac{\delta^2
\sqrt{-\delta}}{\gamma^2}\int_{-\infty}^{\infty}
\xi^2\left(1-\xi^2\right) \left( 1-\frac{9}{4}\frac{\delta^2}{\gamma^2} \left(\xi^2 -1 \right)^2 \right)
{\rm d} \xi.
\end{equation}

Finally the result of integration one has a following expression:
\begin{equation}
\label{eqB.4}   
I_{2} = \frac{\delta^2 \sqrt{-\delta}}{\gamma^2} \left(
-\frac{6}{5}+\frac{72 \delta^2}{70 \gamma^2}\right). \nonumber
\end{equation}

On the the hand the integral $I_1$ can be written as follows

\begin{eqnarray}
\label{eqB.5}
I_1(t_0) &=& \int_{-\infty}^{\infty}  v^*(t)x^*(t) \cos (2 
\omega(t+t_0)) {\rm d} t \\
&=& - \frac{9~\delta^2}{2 \gamma^2} \int_{-\infty}^{\infty} \frac{ 
t~\sqrt{-\delta}}{2}~
\frac{\tanh\left(\frac{\sqrt{-\delta}t}{2}\right)}{\cosh^2 
\left(\frac{\sqrt{-\delta}t}{2}\right)} \left[ 1- \tanh^2 
\left(\frac{\sqrt{-\delta}t}{2}\right) \right] \cos (2 \omega(t+t_0)) {\rm d} t. 
\nonumber 
\end{eqnarray}
what can be expressed further in terms of more elementary integrals $I_1^A$ and $I_1^B$
\begin{equation}
\label{eqB.6}
I_1(t_0)= - \frac{3\delta^2}{2 \gamma^2} (I_1^A - 3 I_1^B) \sin(2\omega t_0),
\end{equation}
which are given by:
\begin{eqnarray}
\label{eqB.7}
I_1^A &=& \int_{-\infty}^{\infty}  \frac{\tanh \tau}{\cosh^2 
\tau} \sin \left( \frac{4 \omega \tau}{\sqrt{-\delta}} \right)  {\rm d} \tau \\
\label{eqB.8}
I_1^B &=& \int_{-\infty}^{\infty}  \frac{\tanh^3 
\tau}{\cosh^2
\tau} \sin \left( \frac{4 \omega \tau}{\sqrt{-\delta}} \right)  {\rm d} \tau.
\end{eqnarray}  
Note that the second integral $I_1^B$ can be easily obtained 
from the first one  $I_1^A$ through integration by parts 
\begin{equation}
\label{eqB.9}
I_1^B= \frac{ \omega'^2}{12} \left( \frac{8}{\omega'^2} -1 \right) I_1^A, 
\end{equation}
where 
\begin{equation}
\label{eqB.10}
\omega'=\frac{4 \omega}{\sqrt -\delta}
\end{equation}
While  $I_1^A$  should be calculated using the residue theorem
\begin{equation}
\label{eqB.11}
\oint f(z) {\rm d} z = 2 \pi {\rm i} \sum_{k=1}^N 
{\rm Res}[f(z),z_k],
\end{equation}
where
\begin{equation}
\label{eqB.12}
{\rm Res}[f(z),z_k]= \frac{1}{(m-1)!} \lim_{z \rightarrow z_k} 
\frac{{\rm d}^{ m-1}}{{\rm d} z^{m-1}} \left[(z-z_k)^mf(z)\right].
\end{equation}

In our case 
\begin{equation}
\label{eqB.13}
f(z)=4 \frac{\exp (z) - \exp( -z)}{(\exp (z) + \exp( -z))^3} \exp 
\frac{{\rm i} 4 \omega z }{\sqrt{-\delta}} 
\end{equation}
 where on the real axis (Fig. B.1) ${\rm Re} z=\tau$:
\begin{equation}
\label{eqB.14}
{\rm Im} f(z)= \frac{\tanh \tau}{\cosh^2
\tau} \sin \left( \frac{4 \omega \tau}{\sqrt{-\delta}} \right).
\end{equation}
The multiplicity of each pole of the complex function $f(z)$ (Eq. 
\ref{eqB.13}): 
\begin{equation}
\label{eqB.15}
z_k=\left( \frac{\pi}{2} + \pi k \right) {\rm i} ~~~{\rm for}~~~k=1,2,3,...
\end{equation}
can be easily determined as $m=3$. After summation of all poles 
(Fig. B.1) we get:

\begin{equation}
\label{eqB.16}
I_1^A=  \frac{8 \pi \omega ^2}{\delta} 
\frac{ 
\sin \left(  \frac{4 \omega z_0}{\sqrt{-\delta}} \right) 
}
{\sinh 
\left(\frac{2 \omega \pi}{\sqrt{-\delta}} \right)}.
\end{equation}

The result of the above analysis can be written in a compact way:

\begin{equation}
\label{eqB.17}   
I_1(t_0) = \frac{12 \pi \delta^2}{\gamma^2} \frac{\omega^2 \left(1 + 4
\omega^2
\right)}{\sinh ( 2 \omega \pi)} \sin (2 \omega t_0).
\end{equation}

\end{document}